\documentstyle[aaspp4]{article}
\begin{document}
 
\lefthead{Elmegreen}
\righthead{Hierarchical Structure in 2D MHD Simulations}
\slugcomment{to appear in ApJ, Vol 527, Dec. 10, 1999}
 
\title{Formation and loss of hierarchical structure in two-dimensional
MHD simulations of wave-driven turbulence in interstellar clouds}

\author{ Bruce G. Elmegreen\altaffilmark{1}}
\altaffiltext{1}{IBM Research Division, T.J. Watson Research Center,
P.O. Box 218, Yorktown Heights, NY 10598, USA, bge@watson.ibm.com}

\begin{abstract} 

Two dimensional compressible magneto-hydrodynamical (MHD) simulations
run for $\sim20$ crossing times on a $800\times640$ grid with two stable
thermal states show persistent hierarchical density structures and
Kolmogorov turbulent motions in the interaction zone between incoming
non-linear Alfv\'en waves. These structures and motions are similar to
what is commonly observed in weakly self-gravitating interstellar
clouds, suggesting that these clouds get their fractal structures from
non-linear magnetic waves generated in the intercloud medium; no
internal source of turbulent energy is necessary. The clumps in the
simulated clouds are slightly warmer than the interclump medium as a
result of magnetic dissipational and compressive heating when the clumps
form. Thus the interclump medium has a lower pressure than the clumps,
demonstrating that the clumps owe their existence entirely to transient
compressive motions, not pressure confinement by the interclump medium.
Clump lifetimes increase with size, and are about one sound
crossing time. 

Two experiments with this model illustrate a possible trigger for star
formation during spontaneous cloud evolution driven by self-gravity and
increased self-shielding. A first test is of the hypothesis that a low
ionization fraction and enhanced magnetic diffusion lead to the
disappearance of clumps smaller than an Alfv\'en wavelength. Two
identical models are run that differ only in the magnetic diffusion
rate. The results show a significant decrease in the magnetic wave
amplitude as the diffusion rate increases, in agreement with
expectations for wave damping, but there is virtually no change in the
density structure or amplitude of the density fluctuations as a result
of this increased diffusion. This is because all of the density
fluctuations are essentially sonic in nature, driven by the noise from
Alfv\'en wave motions outside and at the surface of the cloud. These
sonic disturbances travel throughout the cloud parallel to the mean
field orientation and are not affected by the local magnetic wave
dissipation rate. This result implies that low ionization fractions in
molecular clouds do not necessarily lead to increased cloud smoothing.

The second experiment tests the hypothesis that enhanced density alone
in a self-gravitating cloud leads to wave self-shielding and loss of
incident turbulent energy. Three models with identical conditions
except for the presence or lack of an imposed plane-parallel
gravitational field confirm that externally generated magnetic waves
tend to be excluded from the densest regions of self-gravitating
clouds, and as a result these clouds show a significant loss of density
substructure. This loss of turbulent energy and density substructure
may trigger star formation in the relatively quiescent gas pools that
contain a thermal Jeans mass or more. Such a model fits well with
the hypothesis that the stellar initial mass function comes from the
structure of turbulent hierarchical clouds.  \end{abstract}

Subject Headings: ISM: clouds --- ISM: kinematics and dynamics --- 
ISM: structure --- MHD --- stars: formation --- turbulence

\section{Introduction}

The hierarchical or fractal nature of interstellar clouds has been
proposed to result from turbulence because of the expected compression
from pervasive supersonic flows, and because of the similarity between
cloud structure and the morphology of laboratory turbulence (von
Weizsacker 1951; Sasao 1973; Dickman 1985; Scalo 1985, 1987, 1990; Falgarone
1989; Falgarone \& Phillips 1990; Falgarone, Phillips, \& Walker 1991;
Stutzki, et al. 1991; Mandelbrot 1983; Sreenivasan 1991). 

There are several important differences between interstellar and
laboratory fluids, however (Scalo 1987; Elmegreen 1993a). The
interstellar structure comes from variations in the total gas density,
while the laboratory structure is usually in some tracer of fluid
motions, such as smoke particles or droplets, with the background
density nearly uniform. The interstellar case is also highly magnetic,
with the likely inhibition of some torsional motions from magnetic
tension. Laboratory turbulence is highly vortical. Thus, the analogy
between interstellar and laboratory turbulence is not perfect, even
though the resulting structures are somewhat similar.

The purpose of this paper is to demonstrate that interacting non-linear
magnetic waves provide a physical mechanism for interstellar turbulence
that generates whole clouds and the fractal structures inside of clouds
without internal sources or any special (e.g., power-law) initial
conditions. Computer simulations in two dimensions show cloud and clump
formation at a rapidly cooled, compressed interface between incoming
streams of shear Alfv\'en waves. The resulting structures have power-law
characteristics in both space and velocity. Goldreich \& Sridhar (1995)
also considered turbulence driven by shear Alfv\'en wave interactions,
but treated only the incompressible case. Our results imply that
molecular cloud turbulence and clumpy structures do not depend on star
formation for their generation. They can arise instead from a variety of
energy sources outside the cloud, and get transmitted to the region
where the cloud forms along mildly non-linear magnetic waves. 

The results also imply that a magnetic field is important for general
interstellar turbulence: it distributes the energy from stars and other
sources over large regions at supersonic speeds, and it converts this
energy via non-linear wave mixing into clouds and clumps spanning a wide
range of scales. Intercloud wave damping should not prevent this energy
redistribution because the intercloud wave speed is fast and the energy
sources are relatively infrequent, making the intercloud wavelength
long, perhaps many tens of parsecs. Non-linear damping tends to
occur on the scale of a wavelength when the perturbation speed is
comparable to the Alfv\'en speed (Zweibel \& Josafatsson 1983; Goldreich
\& Sridhar 1995). It occurs on longer scales when the perturbation
speeds are sub-Alfv\'enic, as is the case for the models here.

The magnetic field is more important for simulations of interstellar
turbulence that use a time-dependent energy equation, as is the case
here, than it is for simulations that use a fixed adiabatic index to
relate pressure and density. The fixed index has an artificial energy
source upon decompression that is not present when the full energy
equation is used. As a result, waves and turbulent energy can travel
great distances without needing a magnetic field when there is a fixed
adiabatic index, but they can hardly disperse at all in the field-free
case when the full energy equation is used 
(Elmegreen 1997b, hereafter Paper I). 

In a second part of this paper, we experiment with the hierarchical
structures that result from turbulence to try to understand how diffuse
and translucent clouds, which always seem to have tiny clumps down to
substellar scales, make the conversion to star-forming clouds that have
stellar-mass dense cores. Of course, self-gravity is ultimately
important for this conversion, but the presence of sub-stellar clumps
in all diffuse clouds implies that as long as turbulence is strong, the
stellar-mass clumps that form can always be sheared and sub-divided
into even smaller pieces, preventing self-gravity from dominating
turbulent pressures on stellar scales (e.g., Padoan 1995). This effect
of turbulence is most clearly revealed by the systematic decrease in
the ratio of the clump mass to the turbulent Jeans mass on decreasing
scales in molecular clouds (Bertoldi \& McKee 1992; Falgarone, Puget,
\& P\'erault 1992; Vazquez-Semadeni, Ballesteros-Paredes, \& Rodriguez
1997). Such a decrease follows directly from the Larson (1981) scaling
laws (see Elmegreen 1998). How can stars ever form in a turbulent
medium if non-linear motions always break up the structures into
smaller, more weakly self-gravitating pieces? The answer must be that
star formation begins only after the sub-stellar pieces lose their
turbulence and smooth out into larger mass clumps.

One implication of this model is that fragmentation is not the key to
star formation: star-forming clouds are already fragmented as a remnant
of their pre-star formation turbulence. The {\it key to star formation
is the smoothing of these fragments into stagnant pools larger than a
thermal Jeans mass} (Elmegreen 1999). In an extended, well-connected,
magnetic medium, turbulent dissipation alone is not enough to initiate
this process, because there is a constant flux of turbulent energy into
any particular region from waves and motions on larger scales. Only when
this external flux stops can the agitated motions begin to decay
locally, and only after this decay can the density substructure begin to
disappear, allowing stars to form at the thermal Jeans mass and larger.

With this star formation model in mind, we ran several simulations that
experiment with possible causes of increased cloud smoothing. One lowers
the ionization fraction and thereby decouples the magnetic waves from
the gas (Mouschovias 1991; Myers 1998). Another introduces a density
gradient from a fixed gravitational potential in order to exclude
externally generated waves. The results of these experiments are
discussed in Sections \ref{sect:mag} and \ref{sect:grav}. It seems that
{\it the mere condensation of a cloud resulting from bulk self-gravity is
enough to initiate star formation in Jeans-mass or larger pieces by excluding
external non-linear waves.}

Aside from these applications to star formation, the primary goal of the
present work is to demonstrate that interstellar clouds and their
hierarchically clumpy substructures can arise as transient objects in
the converging parts of supersonic turbulent flows. The general concept
that supersonic turbulence can produce density structure has been
around for a long time, but specific applications to interstellar
clouds were not taken seriously until recently. The early work by Sasao
(1973) on this topic was overwhelmed by the more obvious notion that
HII regions (Hills 1972; Bania \& Lyon 1980), supernovae (Cox \& Smith
1974; Salpeter 1976; McKee \& Ostriker 1977) and combinations of these
pressures (Tenorio-Tagle \& Bodenheimer 1988) directly make clouds
behind moving shock fronts. Indeed there is a lot of evidence for the
bubble structures that are expected from these centralized pressure
sources (Brand \& Zealey 1975). In addition, thermal instabilities
(Field, Goldsmith \& Habing 1969) were proposed to make diffuse clouds,
while magnetic and gravitational instabilities (Parker 1966; Goldreich
\& Lynden-Bell 1965) were proposed to make giant clouds, particularly
the kpc-size condensations in spiral arms (Elmegreen 1979). Everything
in between was supposed to be made by the build-up and dispersal cycle
of collisional agglomeration (e.g., Field \& Saslaw 1965; Kwan 1979).

The importance of turbulence as a structuring agent has now reemerged in
the astronomical literature. This change slowly followed several key
observations, including the recognition of correlated motions in
molecular clouds (Larson 1981) and of pervasive small scale structure in
diffuse (Low et al. 1984; Scalo 1990), ionized (Spangler 1998), and
molecular clouds (e.g., Falgarone, Phillips, \& Walker 1991; Stutzki et
al. 1998; Falgarone et al. 1998). Correlated motions and power-law
structures also appear in HI surveys (Green 1993; Lazarian 1995;
Stanimirovic et al. 1999; Lazarian \& Pogosyan 1999).  LaRosa, Shore
\& Magnani (1999) found correlated motions in a translucent cloud and
suggested the driving source was outside, as in the present model. 

Astrophysical models of cloud
formation also reflected this change by including turbulence
compression as one of several mechanisms (see reviews in Elmegreen
1993a, 1996). Star formation theory changed as well, e.g., by
considering not only the influence of converging turbulent flows on the
gravitational stability of clumps (Hunter \& Fleck 1982), but also the
stability of clumps that specifically formed by this compression and
whose lifetimes were consequently very short (Elmegreen 1993b). Recent
emphasis on cloud formation in turbulence-induced flows is in
Ballesteros-Paredes, Vazquez-Semadeni \& Scalo (1999), with an
application to the Taurus clouds by Hartmann, Ballesteros-Paredes, \&
Vazquez-Semadeni, et al. (1999).  The high latitude clouds (Magnani,
Blitz, \& Mundy 1995) may be examples as well, because of their short
lives.  A generalization of turbulence structures to include both cloud
and intercloud media is in Elmegreen (1997a).

Other astrophysical turbulence simulations have addressed different
problems. In a series of papers, Vazquez-Semadeni, Passot \& Pouquet
(1995, 1996), Passot, Vazquez-Semadeni,\& Pouquet (1995), and
Vazquez-Semadeni, Ballesteros-Paredes, \& Rodriguez (1997) studied the
properties of clouds and star formation in a turbulence model with star
formation heating, gaseous gravity, rotation, magnetism, and other
effects, to simulate a galactic-scale piece of the interstellar medium. 

Gammie, \& Ostriker (1996) used a 1D simulation to show that forces
parallel to the mean field can support a cloud against self-gravity.
Stone, Ostriker, \& Gammie (1998) and MacLow, et al. (1998) found the
rate of energy dissipation in 3D supersonic MHD turbulence to be about
the same as in turbulence without magnetic fields, and suggested that
energy input from embedded stars or other sources was necessary to
maintain cloud structure for more than a crossing time. Vazquez-Semadeni
(1994), Nordlund, \& Padoan (1998), and Scalo et al. (1998) found a
log-normal
probability density distribution generated by compressible turbulence. 
Hierarchical turbulent structures like those produced here have also
been generated in the wakes behind shocked clouds (Klein, McKee \&
Colella 1994; Mac Low et al. 1994; Xu \& Stone 1995). 

Padoan (1998) proposed that interstellar turbulence is super-Alfv\'enic
in order to get significant compression. This is different from the
cases considered here, where all the cloud motions are sub-Alfv\'enic.
We believe his application of MHD results to interstellar turbulence is
inappropriate because the interstellar structure is hierarchically
clumped, and super-Alfv\'enic turbulence can only be super-Alfv\'enic
for one level in a hierarchy of structures. Once a region is compressed,
it becomes sub-Alfv\'enic because the magnetic pressure increases much
more than the thermal. Thus the process of turbulence compression would
stop after only one level in his model and could never get the observed
hierarchical structures. In the MHD simulations discussed here, the
hierarchical structure comes from sonic and slightly supersonic motions
along the field, driven by transverse magnetic waves. Such sonic
disturbances can divide the gas into very small pieces regardless of the
local field strength. 

The present work in two-dimensions is an extension of the same cloud
formation problem considered in one-dimension earlier, first with the
simple demonstration that non-linear transverse Alfv\'en waves push the
gas along and make high density structures when they interact (Elmegreen
1990), and then with randomly driven transverse waves at the ends of a
1D grid making hierarchical density structures in the remote regions
between (Elmegreen 1991, and Paper I). The first 2D simulations of
clumpy structure from interacting magnetic waves were in Yue et al.
(1993).

This 1D compression between wave sources is analogous to the ``turbulent
cooling flow'' discussed by Myers \& Lazarian (1998), but it is not
exactly the same. Here a dense region forms because of a convergence and
high pressure from external magnetic waves and the thermal cooling that
follows at the compressed interface.  Turbulent dissipation at the
interface is not as important as thermal cooling for this density
enhancement. The turbulent cooling envisioned by Myers \&
Lazarian leads to a macroscopic thermal instability (Struck-Marcell
\& Scalo 1984; Tomisaka 1987; Elmegreen 1989). If this were to operate
in addition to the thermal variations modeled here, then the density
enhancement in the converging region would be even larger that what
thermal effects alone give.  However, the turbulent energy in the cloud
is continuously resupplied from outside, and it gets in
easily until the cloud density contrast becomes large (Sect. \ref{sect:grav}.
As a result, the internal turbulent energy does not decrease
much for small perturbations in the density, and there is no
runaway turbulent cooling in our models.

In what follows, we discuss the MHD algorithm and tests
of the numerical code in Section \ref{sect:intro}.  This method of
solving fluid equations is new to the
astronomy community, so we summarize the essential points in some
detail (it was used also in Paper I for 1D turbulence, but the
equations were not given there). In section \ref{sect:hierarchical}, 
the basic MHD simulation that generates hierarchical and scale-free 
structure from turbulence is discussed.  Models with enhanced magnetic
diffusion are in section \ref{sect:mag} and models with imposed,
plane-parallel gravity are in section \ref{sect:grav}.

\section{The relaxation algorithm of Jin \& Xin}
\subsection{Introduction}
\label{sect:intro}

The relaxation algorithm developed by Jin \& Xin (1995) has been adapted here
to include magnetic fields.  We also added heating and cooling rates to the
energy equation to give two
stable thermal states.  This is analogous to what we did 
for the one dimensional problem (Paper I).

The Jin \& Zin method is a way to solve
systems of conservation equations with no need for artificial viscosity in the
equation of motion.   It does this by
dividing the primary equations of motion and continuity into three equations,
\begin{eqnarray}
{{\partial {\bf S}}\over{\partial t}}+{{\partial {\bf v}}\over
{\partial x}} +
{{\partial {\bf w}}\over{\partial y}}=0,\label{eq:basic1}\\
{{\partial {\bf v} }\over{\partial t}}+A{{\partial {\bf S} }\over{\partial x}}=
-{1\over{\epsilon}}
\left({\bf v}-{\bf F}_x\left[{\bf S}\right]\right)\\
{{\partial {\bf w}}\over{\partial t}}+B{{\partial {\bf S}}\over{\partial y}}=
-{1\over{\epsilon}}
\left({\bf w}-{\bf F}_y\left[{\bf S}\right]\right).
\label{eq:basic}
\end{eqnarray}
The vector {\bf S} is a vector of physical variables consisting of density, 
momentum density, magnetic field strength, and energy density, as written below. The
vectors ${\bf F}_x$ and ${\bf F}_y$ are the two spatial components of the total
forces on each physical variable, 
which are, in general, functions of the physical variables.
The scalar $\epsilon$ is a relaxation rate, taken to be a small positive
parameter with the dimensions of time, 
and $A$ and $B$ are diagonal matrices with the dimensions of velocity squared.
Vectors {\bf v} and {\bf w} are intermediate variables that ``relax'' to the
force vectors, ${\bf F}_x$ and ${\bf F}_y$, respectively, on the time scale
$\epsilon$.

To ensure stability, $A>\left(\partial {\bf F}_x\left[{\bf S}\right]/\partial 
{\bf S}\right)^2$
and  $B>\left(\partial {\bf F}_y\left[{\bf S}\right]/\partial 
{\bf S}\right)^2$
for all {\bf S}, which means that each
element of $A$ and $B$ has to equal or exceed the square of the maximum velocity that 
occurs in the simulation.  For our calculations, we take a constant $a\equiv A^{1/2}=B^{1/2}=$
several times the Alfv\'en speed in the unperturbed gas.  

The Jin \& Zin method requires that the physical equations
for the problem be written
in conservative form (see also Dai \& Woodward 1998). 
It also prescribes how to discretize the spatial and time
steps in an efficient and stable way. 

\subsection{Spatial Discretization}
\label{sect:spatial}

The spatial discretization is done in the manner suggested by
Jin \& Xin (1995).  This is second-order accurate, and follows from
van Leer (1979), using a piecewise linear interpolation.  
The resulting discrete forms of equations \ref{eq:basic1}--\ref{eq:basic} are, for
each component of the vectors {\bf S}, {\bf v}, ${\bf F}_x$, and ${\bf F}_y$ :
\begin{eqnarray}
{{\partial}\over{\partial t}} S + D_x\left( v,S\right)+
D_y\left(w,S\right)=0\\
{{\partial}\over{\partial t}} v + a^2D_x( v,S)=
-{1\over{\epsilon}}\left(v-F_x\left[S\right]\right),\\
{{\partial}\over{\partial t}} w + a^2D_y( w,S)=
-{1\over{\epsilon}}\left(w-F_y\left[S\right]\right),
\end{eqnarray}
where the derivatives for $v$ and $w$ are
\begin{eqnarray}
D_x\left(v_{i,j},S_{i,j}\right)={{1}\over{2\Delta}}
\left(v_{i+1,j}-v_{i-1,j}\right)-
{{a}\over{2\Delta}}\left(S_{i+1,j}-2S_{i,j}+S_{i-1,j}\right)\nonumber\\
-{{1}\over{4}}\left(\sigma_{x;i+1,j}^- -\sigma_{x;i,j}^- -\sigma_{x;i,j}^+
+\sigma_{x;i-1,j}^+\right),\\
D_y\left(w_{i,j},S_{i,j}\right)={{1}\over{2\Delta}}
\left(w_{i,j+1}-w_{i,j-1}\right)-
{{a}\over{2\Delta}}\left(S_{i,j+1}-2S_{i,j}+S_{i,j-1}\right)\nonumber\\
-{{1}\over{4}}\left(\sigma_{y;i,j+1}^- -\sigma_{y;i,j}^- -\sigma_{y;i,j}^+
+\sigma_{y;i,j-1}^+\right).
\end{eqnarray}
The indices $(i,j)$ represent the $(x,y)$ positions in the 
computational grid,
$\Delta$ is a constant spatial step size, $a$ is the maximum velocity 
for the problem, and $\sigma$ is a derivative correction, given by
\begin{eqnarray}
\sigma_{x;i,j}^{\pm}={{1}\over{\Delta}}\left(v_{i+1,j}\pm
aS_{i+1,j}-\left[v_{i,j}\pm aS_{i,j}\right]\right)\phi\left(\theta_{x;i,j}^\pm\right)\\
\sigma_{y;i,j}^{\pm}={{1}\over{\Delta}}\left(w_{i,j+1}\pm
aS_{i,j+1}-\left[w_{i,j}\pm aS_{i,j}\right]\right)\phi\left(\theta_{y;i,j}^\pm\right)
\end{eqnarray}
where
\begin{eqnarray}
\theta_{x;i,j}^\pm={{v_{i,j}\pm aS_{i,j} -\left[v_{i-1,j}\pm aS_{i-1,j}\right]}\over
{v_{i+1,j}\pm aS_{i+1,j} -\left[v_{i,j}\pm aS_{i,j}\right]}},\\
\theta_{y;i,j}^\pm={{w_{i,j}\pm aS_{i,j} -\left[w_{i,j-1}\pm aS_{i,j-1}\right]}\over
{w_{i,j+1}\pm aS_{i,j+1} -\left[w_{i,j}\pm aS_{i,j}\right]}},
\end{eqnarray}
and 
\begin{eqnarray}
\phi\left(\theta\right)={\rm max}\left(0,{\rm min}\left[1,\theta\right]\right)
\end{eqnarray}
is a relatively smooth, slope-limiting function.

\subsection{Time Discretization}
\label{sect:t}

The time stepping is done with
a second-order, total variation diminishing (TVD),
Runge-Kutta splitting time discretization that keeps the
convection terms explicit and the lower order terms implicit
(Jin 1995). 
We write this discretization method in terms of the current time step, with
superscript $n$, the next time step, with index $n+1$, and
intermediate values with superscripts $*$, $**$, (1), and (2) 
(see also Jin \& Xin 1995):
\begin{eqnarray}
{\bf S}^{*}={\bf S}^n\\
{\bf v}^{*}={{{\bf v}^n-{\bf F}_x\left({\bf S}^{*}\right)
dt/\epsilon}\over{1-dt/\epsilon}}\\
{\bf w}^{*}={{{\bf w}^n-{\bf F}_y\left({\bf S}^{*}\right)
dt/\epsilon}\over{1-dt/\epsilon}}\\
{\bf S}^{(1)}={\bf S}^{*}-D_x\left({\bf v}^{*},{\bf S}^{*}\right)dt
-D_y\left({\bf w}^{*},{\bf S}^{*}\right)dt\\
{\bf v}^{(1)}={\bf v}^{*}-a^2 D_x\left({\bf v}^{*},{\bf S}^{*}\right)dt\\
{\bf w}^{(1)}={\bf w}^{*}-a^2 D_y\left({\bf w}^{*},{\bf S}^{*}\right)dt\\
{\bf S}^{**}={\bf S}^{(1)}\\
{\bf v}^{**}={{{\bf v}^{(1)}+{\bf F}_x\left({\bf S}^{**}\right)
dt/\epsilon-2\left[{\bf v}^{*}-{\bf F}_x\left({\bf S}^{*}\right)\right]dt/\epsilon}
\over{1+dt/\epsilon}}\\
{\bf w}^{**}={{{\bf w}^{(1)}+{\bf F}_y\left({\bf S}^{**}\right)
dt/\epsilon-2\left[{\bf w}^{*}-{\bf F}_y\left({\bf S}^{*}\right)\right]dt/\epsilon}
\over{1+dt/\epsilon}}\\
{\bf S}^{(2)}={\bf S}^{**}-D_x\left({\bf v}^{**},{\bf S}^{**}\right)dt
-D_y\left({\bf w}^{**},{\bf S}^{**}\right)dt\\
{\bf v}^{(2)}={\bf v}^{**}-a^2 D_x\left({\bf v}^{**},{\bf S}^{**}\right)dt\\
{\bf w}^{(2)}={\bf w}^{**}-a^2 D_y\left({\bf w}^{**},{\bf S}^{**}\right)dt\\
{\bf S}^{n+1}=\left({\bf S}^n+{\bf S}^{(2)}\right)/2\\
{\bf v}^{n+1}=\left({\bf v}^n+{\bf v}^{(2)}\right)/2\\
{\bf w}^{n+1}=\left({\bf w}^n+{\bf w}^{(2)}\right)/2.
\end{eqnarray}
The time step has to satisfy $dt>>\epsilon$ to make the ${\bf v}^*$
and ${\bf w}^*$ equations stable; we use $dt=10^{-2}$ and $\epsilon=10^{-6}$
for the standard and gravitating slab cases, and $dt=10^{-3}$ and $\epsilon=10^{-9}$
for the experiment with low magnetic diffusion. (Reasons for these timesteps
are given in sections \ref{sect:disc} and \ref{sect:mag}.)

\subsection{Physical Equations in Conservative Form}
\subsubsection{Mass Conservation}

The variables ${\bf S}$, and the force vectors ${\bf F}_x$ and 
${\bf F}_y$, come from the
physical equations written in conservative form, which is 
$\partial {\bf S}/\partial t + \nabla\cdot {\bf F}=0$. The 
mass conservation equation,
\begin{equation}                                                                
{{\partial \rho} \over{\partial t}} + \nabla\cdot\rho {\bf v} = 0             
\end{equation}
is already in this form for mass density $\rho$ and
velocity vector {\bf v}. We rewrite this and other equations 
in terms of indices $j$ to
designate the spatial vector components $x$ and $y$, 
\begin{equation}                                                                
{{\partial \rho} \over{\partial t}} +                                           
{{\partial \rho v_j} \over{\partial x_j}}.                                      
\end{equation}  
Then the first terms in the ${\bf S}$, ${\bf F}_x$ and ${\bf F}_y$ vectors are 
\begin{equation}
S_1=\rho \;\;;\;\; F^j_1=\rho v_j
\end{equation}
where $j=1$ for the $x$ component and $j=2$ for the $y$ component.

\subsubsection{Equation of Motion}

The equation of motion can be written in 
the same way, but gravity, viscosity, and perhaps other forces add additional 
terms to the right hand side:
\begin{equation}
{{\partial {\bf S}}\over{\partial t}}+\nabla\cdot{\bf F}={\bf D}.
\end{equation}
The viscous terms and gravity are inside {\bf D}.  We do not include
any viscosity in the equations because molecular viscosity is
negligible in interstellar clouds; all of the energy dissipation
occurs in the cooling term, $\Lambda$, which appears in the energy equation.
We also ignore self-gravity for simplicity, but some runs have
fixed gravity to investigate clumpy structure in regions
with smooth density gradients. 

To include magnetic diffusion, we write separate equations
of motion for the neutrals and ions. For the ions (distinguished
by the symbol "+"), the equation of motion is
\begin{equation}                                                                
\rho_+\left({{\partial {\bf v}_+}\over{\partial t}}                                        
+{\bf v}_+\cdot\nabla{\bf v}_+\right)                                                
=-\nabla P_+ +\rho_+ {\bf g} + {1\over{4\pi}}{\bf B}\cdot\nabla{\bf B}                        
- {1\over{8\pi}}\nabla{\bf B}\cdot{\bf B}                                     
- n_+ <\sigma a_t> \mu ({\bf v}_+-{\bf v})n                                                   
\end{equation} 
where $<\sigma a_t>$ is the ion-neutral thermal collision rate for thermal
speed $a_t$, $\mu$ is the reduced ion-neutral mass, $n_+$ is the charged
particle density, $n$ is the total particle density,
{\bf B} is the magnetic field strength, and {\bf g} is the gravitational
acceleration.  The total mass density of the
ions is negligible in interstellar clouds, so the inertial terms can
be dropped. Then we get
\begin{equation}
{1\over{4\pi}}{\bf B}\cdot\nabla{\bf B}
- {1\over{8\pi}}\nabla{\bf B}\cdot{\bf B}
= n_+ <\sigma a_t> \mu n({\bf v}_+-{\bf v})
\equiv \omega_+ \rho ({\bf v}_+-{\bf v}),
\label{eq:vdiff}
\end{equation}
where $\omega_+$ is defined as a collision rate per unit volume, corrected
for reduced mass.

For the neutral particles, the equation of motion is
\begin{equation}                                                                
\rho\left({{\partial {\bf v}}\over{\partial t}}                                
+{\bf v}\cdot\nabla{\bf v}\right)                                             
=-\nabla P +\rho {\bf g} -\omega_+ \rho ({\bf v}-{\bf v}_+).
\label{eq:motion}
\end{equation} 
For this equation, $-\omega_+ \rho (v-v_+)$ may be substituted from above to give
\begin{equation}                                                                
\rho\left({{\partial {\bf v}}\over{\partial t}}                                
+{\bf v}\cdot\nabla{\bf v}\right)                                             
=-\nabla P+\rho {\bf g} 
+{1\over{4\pi}}{\bf B}\cdot\nabla{\bf B}                                      
- {1\over{8\pi}}\nabla{\bf B}\cdot{\bf B}.                                   
\end{equation}
The ion equation will be used again to write the
time evolution of the magnetic field in terms of convective and
diffusion terms. 

This neutral equation of motion can be converted to the conservative form:
\begin{equation}                                                                
{{\partial \rho v_i}\over{\partial t}} +                                        
{{\partial \rho v_iv_j}\over{\partial x_j}} +                                   
{{\partial }\over{\partial x_j}} \left[\left(p+{{B^2}\over{8\pi}}\right)        
\delta_{ij}-{{B_iB_j}\over{4\pi}}\right] = \rho g_i                             
\end{equation}
where $\delta_{ij}=1$ for $i=j$.  As a result, the second and third
components of $S$, $F$, and $D$, designated by subscripts $i+1=2,3$, are, for 
spatial coordinate indices $i=1,2$ and $j=1,2$:
\begin{equation}
S_{i+1}=\rho v_i
\label{eq:mom}
\end{equation}
\begin{equation}                                                                
F_{i+1}^j=\rho v_iv_j+                                                          
\left(p+{{B^2}\over{8\pi}}\right)\delta_{ij}-{{B_iB_j}\over{4\pi}}              
\end{equation}                                                                  
\begin{equation}                                                                
D_{i+1}=\rho g_i.\end{equation}
With this notation, the equation of motion is
\begin{equation}                                                                
{{\partial S_{i+1}}\over{\partial t}}+                                          
{{\partial F_{i+1}^j}\over{\partial x_j}}=D_{i+1},
\label{eq:mot}
\end{equation} 
where $j=1$ and 2 for the $x$ and $y$ components of the force matrix, and
$i=1$ and 2 for the $x$ and $y$ components of the momentum density.

\subsubsection{Equation of Magnetic Field Evolution}

Magnetic field evolution
is given by the convection of ions in the usual way:
\begin{equation}   
{{\partial {\bf B}}\over{\partial t}}=\nabla\times({\bf v_+}\times{\bf B}).
\end{equation}  
This may be written in terms of the neutral velocity {\bf v}:
\begin{equation}                                                                
{{\partial {\bf B}}\over{\partial t}}=
\nabla\times\left({\bf v}\times{\bf B}\right)-
\nabla\times\left([{\bf v}-{\bf v_+}]\times{\bf B}\right)
\end{equation}
where ${\bf v}-{\bf v_+}$ comes from equation (\ref{eq:vdiff}).

To write this equation in conservative form, we have to expand
$\nabla\times\left({\bf v}\times{\bf B}\right)$ and rearrange terms.  The result is
\begin{equation}
{{\partial B_i}\over{\partial t}}+{{\partial }\over{\partial x_j}}
\left[v_jB_i-v_iB_j\right]+{{\partial }\over{\partial x_j}}                                                
\left[(v_{j+}-v_j)B_i-(v_{i+}-v_{i})B_j\right]=0.
\end{equation}
The last two terms come from equation (\ref{eq:vdiff}), which may be re-written as
\begin{equation}                                                                
v_{i+}-v_i=                                                                     
{1\over{\omega_+\rho}}                                                          
\left[{1\over{4\pi}}                                                            
{{\partial \left(B_kB_i-0.5\delta_{ik}B^2\right)}\over{\partial x_k}}           
\right].
\label{eq:vdiffij}                                                                    
\end{equation} 

This form of the magnetic field evolution equation 
indicates that the computational variables are, for $i=1,2$,
\begin{equation}
S_{i+4}=B_i,
\end{equation}
\begin{equation}
F_{i+4}^j=v_jB_i-v_iB_j
\end{equation}
and 
\begin{equation}
D_{i+4}=-{{\partial }\over{\partial x_j}}
\left[(v_{j+}-v_j)B_i-(v_{i+}-v_i)B_j\right]
\end{equation}
where equation (\ref{eq:vdiffij}) has to be used as well. Thus the equation of
magnetic field evolution becomes similar to equation (\ref{eq:mot}) for $i=1,2$,
\begin{equation}
{{\partial S_{i+4}}\over{\partial t}}+
{{\partial F_{i+4}^j}\over{\partial x_j}}=D_{i+4}.
\end{equation}

\subsubsection{Energy Equation} 

The energy equation in conventional form is
\begin{equation}
{{\partial U}\over{\partial t}}+{\bf v}
\cdot\nabla U+(U+p)\nabla\cdot{\bf v}=\Gamma-\Lambda,
\end{equation}
where                                                                           
\begin{equation}
U={{p}\over{\gamma-1}}\end{equation}
is the energy density, $p$ is the thermal pressure, $\Gamma$ and
$\Lambda$ are the thermal heating and cooling rates, and $\gamma=5/3$
is the ratio of specific heats for a monatomic or cold molecular gas. 
Note that these are the fundamental
physical variables, and not effective variables, such as turbulent
pressure or effective $\gamma$, which is sometimes used
for shock fronts to automatically include heating and cooling behind the
shock. Here the computer code generates by itself what we
have come to think of as ``turbulent pressure,'' and the heating
and cooling terms determine the relation between temperature and
pressure in a self-consistent way.

The energy equation is converted into conservative form by adding it to
the dot product of the velocity with the equation of motion, and
then substituting from other equations. The result has the same general form
as before, now written with subscript 6 because for a two
dimensional problem, the energy equation is the 6th component of the general
vector equation:
\begin{equation}
{{\partial S_{6}}\over{\partial t}}+
{{\partial F_{6}^j}\over{\partial x_j}}=D_{6}.
\end{equation}
where 
\begin{equation}   
S_{6}=U+0.5\rho v^2+B^2/(8\pi)
\end{equation}
is the total energy,
\begin{equation}
F_6^j=v_j\left[U+p+0.5\rho v^2+B^2/(4\pi)\right]-B_jB_iv_i/(4\pi)
\end{equation}
is the energy flux, and
\begin{equation}
D_8=-{{B_i}\over{4\pi}}{{\partial }\over{\partial x_j}}
\left[X_jB_i-X_iB_j\right]+\Gamma-\Lambda+ \rho v_ig_i,
\end{equation}
where                                                                           
\begin{equation}                                                                
X_j= {1\over{\omega_+\rho}}                                                     
\left[{1\over{4\pi}}                                                            
{{\partial \left(B_kB_j-0.5\delta_{jk}B^2\right)}\over{\partial x_k}}           
\right] .                                                                        
\end{equation} 
Throughout these derivations, we have used the convention that repeated
indices are summed.

The heating and cooling terms are designed to give two stable
temperature states and an unstable state between them. 
This is done by taking
\begin{equation}
\Gamma-\Lambda=-\Gamma_0\rho p C\left(C-0.5\right)
\left(C-1\right)/0.04811
\end{equation}
where $C=\log\left(p/\rho\right)$.
For large temperatures, $C\sim1$, the quantity $\Gamma-\Lambda$
is zero and decreasing with increasing $C$, so the region cools if $C>1$ and
heats up if $C<1$; this is stable behavior with equilibrium at $C=1$. 
For low temperature, 
$C\sim0$, $\Gamma-\Lambda\sim0$ and decreasing again, so 
the region is stable there too. 
At intermediate temperatures, where $C\sim0.5$, $\Gamma-\Lambda$
increases with $C$, so $C>0.5$ leads to more heating and a further
increase in $C$, while $C<0.5$ leads to more cooling and
a further decrease in $C$; this is unstable behavior. The
equilibrium temperatures correspond to $p/\rho=$ 1 and 10. 
For comparison, 
the initial temperature in all of our simulations corresponds
to the stable solution, $p/\rho=1$, and the initial Alfv\'en speed, 
$B/\left(4\pi\rho\right)^{1/2}$, is 10.
The coefficient 0.04811 in $\Gamma-\Lambda$ makes the peak values of the
cubic part, $-C\left(C-0.5\right)\left(C-1\right)$,
equal to $-1$ at $\log p/\rho=0.211$ and 
$+1$ at $\log p/\rho=0.789$.

The constant $\Gamma_0=10^{-3}$ determines the cooling time.
If we write the magnetic field-free
equation of energy as 
$Dp/Dt=\left(\gamma p/\rho\right)D\rho/Dt+\left(\gamma-1\right)
\left(\Gamma-\Lambda\right)$,
then the cooling time is 
$t_{cool}=\left(\left[\gamma-1\right]\Gamma_0\rho\right)^{-1}$.
Initially, this is 
$3/\left(2\Gamma_0\right)$ for $\gamma=5/3$ and $\rho=1$, and this is
1500 for $\Gamma_0=10^{-3}$.   This is about the duration of the
simulations, so the local cooling time is
approximately the simulation run time divided
by the density.  

\subsubsection{Discussion}
\label{sect:disc}

This completes the derivation of the physical equations
in conservative form. The variables actually used by the
computer are the $S$ variables, so all these equations
for $F$ and $D$ have to be written in terms of these
$S$ variables. This means, for example, that $v_1$ in an equation has
to be written as $S_2/S_1$, and so on. The $S$ variables are
initialized after consideration of the physical problem, 
{\bf v} is initialized to ${\bf F}_x$, and {\bf w} is initialized
to ${\bf F}_y$.  Then the $S$ variables are incremented in time, 
as discussed in Section \ref{sect:t}.

The unit of time in the simulation is equal to the crossing time of a
coordinate cell ($dx=dy=1$) at a velocity given by the initial ratio of
$P/\rho=1$. The time step size in the simulation is much smaller than
this time unit, because many types of motions are faster than 1 velocity
unit. Initial Alfv\'en waves move at a speed of 10 velocity units, sound
waves in the warm phase move at a speed of $10^{1/2}\gamma=5.3$, and
Alfv\'en waves in the warm phase, where the density is low
($\rho\sim0.1$) move at $10/\left(\rho\right)^{1/2}\sim30$. Thus we use
a time step of 0.01, which is short enough to follow all of these
motions.

The equations were solved on an IBM SP parallel computer, with
computational space divided up between processors, and communication
between processors done with message passing (MPI). A typical run of
1536 time units, which is 153600 time steps, on a grid of
$800\times640$, with magnetic diffusion, heating and cooling, took about
150 node-days on an IBM SP with the Power2sc processor. We report three
of these runs here, one without gravity and two with a fixed
gravitational potential, plus two other runs on a $400\times320$ grid to
test the importance of magnetic diffusion.

\subsection{Tests}

\subsubsection{Linear Waves}

Several tests were run to check the accuracy of the MHD solutions. 
In all tests, magnetic diffusion was turned off and $\Gamma-\Lambda$ was
set equal to zero.  All simulations used 64-bit floating point accuracy.

In a
simulation with 800 cells and a constant value of the magnetic field strength
and density, a perturbation was given to the velocity in a direction
perpendicular to the field at a grid position halfway between the two
boundaries. The perturbation was very low in
amplitude ($10^{-2}$ times the sound speed), and it had a sinusoidal
time dependence to generate a smooth Alfv\'en wave. The average wavelength of
the wave was measured to be within the factor
$9.6\times10^{-5}$ of the theoretical
prediction (200 cells), indicating that the linearized Alfv\'en speed has this
accuracy. 

\subsubsection{Shock jump conditions}

\begin{figure} 
\vspace{6.in}
\includegraphics{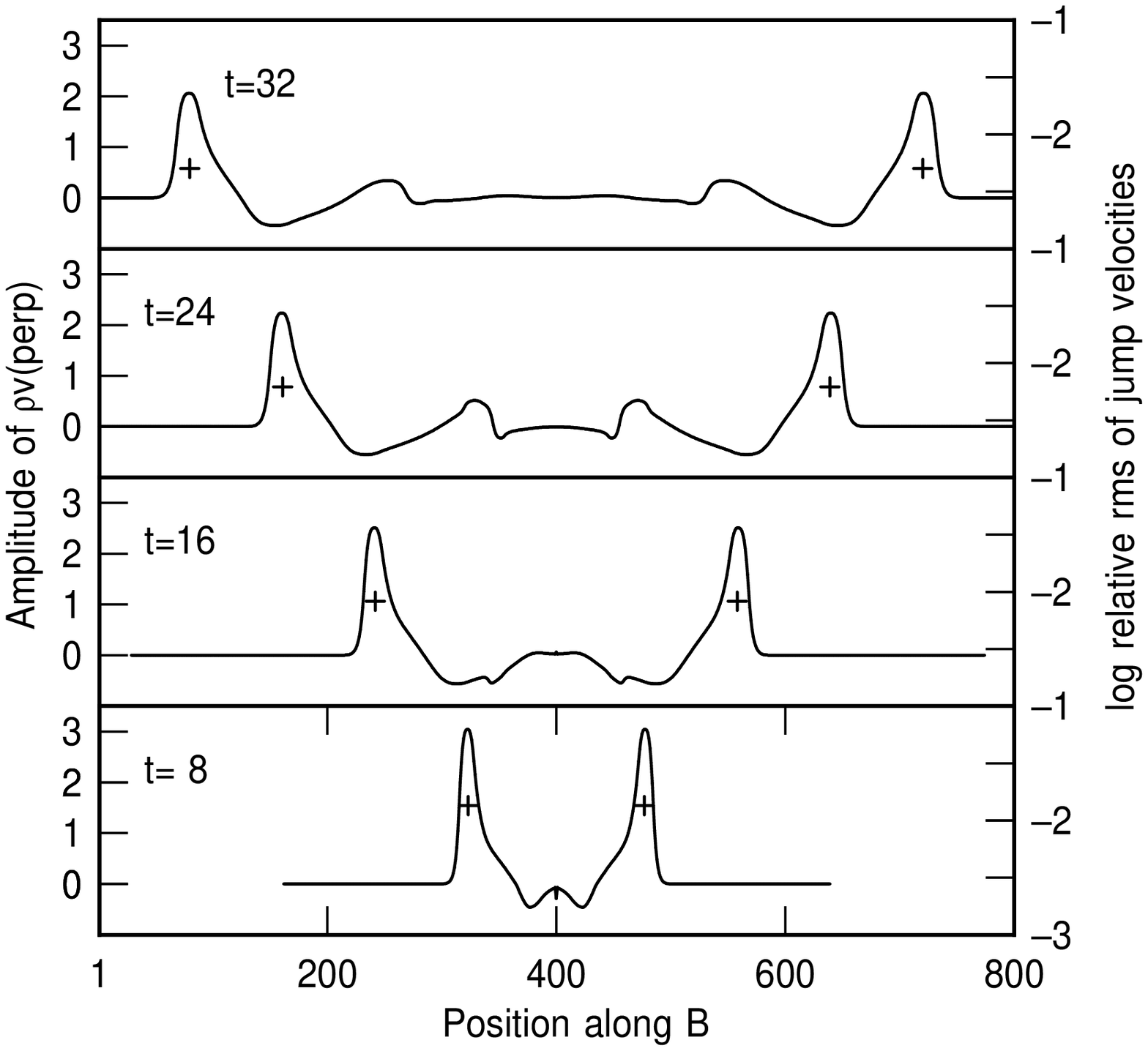}
\caption{The perpendicular momentum flux, $\rho v_x$, is shown as a
function of position along the magnetic field for four time steps
following a time-sinusoidal perturbation in the center. The left-hand
axis shows the amplitude of the momentum. The perturbation drives waves
outward, and these waves steepen and push matter along with them,
slowing and weakening as they go. The relative rms deviations between the
four measures of the shock speed, from the four jump conditions, are
shown by plus signs at the grid points where the density peaks,
using the right-hand axes to indicate the amplitude. The rms deviation
is typically around 1\%. } \label{fig:test} \end{figure}

In another simulation with the same initial conditions, a 
larger perpendicular velocity was applied 
to the same centralized grid point for one sinusoidal cycle,
and the resulting two diverging
waves quickly steepened into shock fronts as they moved away from this point
(cf. Fig. \ref{fig:test}). 
The shocks
continued along the field lines gathering mass and decreasing
in amplitude. The accuracy
of the jump conditions was determined from the rms dispersion
of the various measures of the shock speed: 
\begin{equation}
\xi_1=[m_y]/[\rho],\end{equation}
\begin{equation}
\xi_2={{[m_y^2/\rho]+[p]+[B_x^2]/8\pi}\over{[m_y]}},\end{equation}
\begin{equation}
\xi_3={{[m_xm_y/\rho]-B_x[B_y]/4\pi}\over{[m_x]}},\end{equation} 
and 
\begin{equation}
\xi_4={{[B_xm_y/\rho]-B_y[m_x/\rho]}\over{[B_x]}},\end{equation} 
where $m_y=\rho v_y$ and $m_x=\rho v_x$ are the momentum densities
parallel and perpendicular to the initial field
(equal to $S_3$ and $S_2$ in the notation of Eq. \ref{eq:mom}).
The time evolution of the perpendicular momentum, $m_x$,
is in figure \ref{fig:test}. 

The amplitude of the perturbation was such that the
velocity of the resulting motion was intermediate between the sound
speed and the Alfv\'en speed (1 and 10 in these units, respectively).
This is the regime of the cloud simulations discussed in the rest of
this paper, based on observations of velocities and Alfv\'en speeds in 
the interstellar medium. 

The jump condition expressions given above are the shock speeds from the
mass flux parallel to the field and to the direction of
propagation of the wave, the parallel momentum flux, perpendicular
momentum flux, and perpendicular magnetic field flux, respectively.
These quantities are evaluated at the grid point with the maximum
density behind the shock. The shock jump differences denoted by square brackets
$[...]$ are determined from the differences between the values of various
quantities at the density peak and values at points ahead of the shocks by 20
grid spaces (this number 20 does not matter, as long as it places the
preshock condition in the unperturbed area).  The relative 
rms deviation between all four determinations of the shock velocity
is a measure of the accuracy of the jump conditions. This relative rms
value is shown in Figure \ref{fig:test} as a plus symbol, using the
right hand axis. It is typically 1\% or less, indicating that the
shock jump conditions are satisfied to within this accuracy.  Note that
the shock fronts are not sharp at late times in this test because the
velocity is less than the Alfv\'en speed (compare to Fig. \ref{fig:riemann}
below).

Analogous simulations with perturbations perpendicular to the initial
uniform field gave the same order of magnitude for the accuracy of
the jump conditions. 

\subsubsection{Measurement of the $\nabla\cdot{\bf B}=0$ error}

\begin{figure} 
\vspace{6.in}
\includegraphics{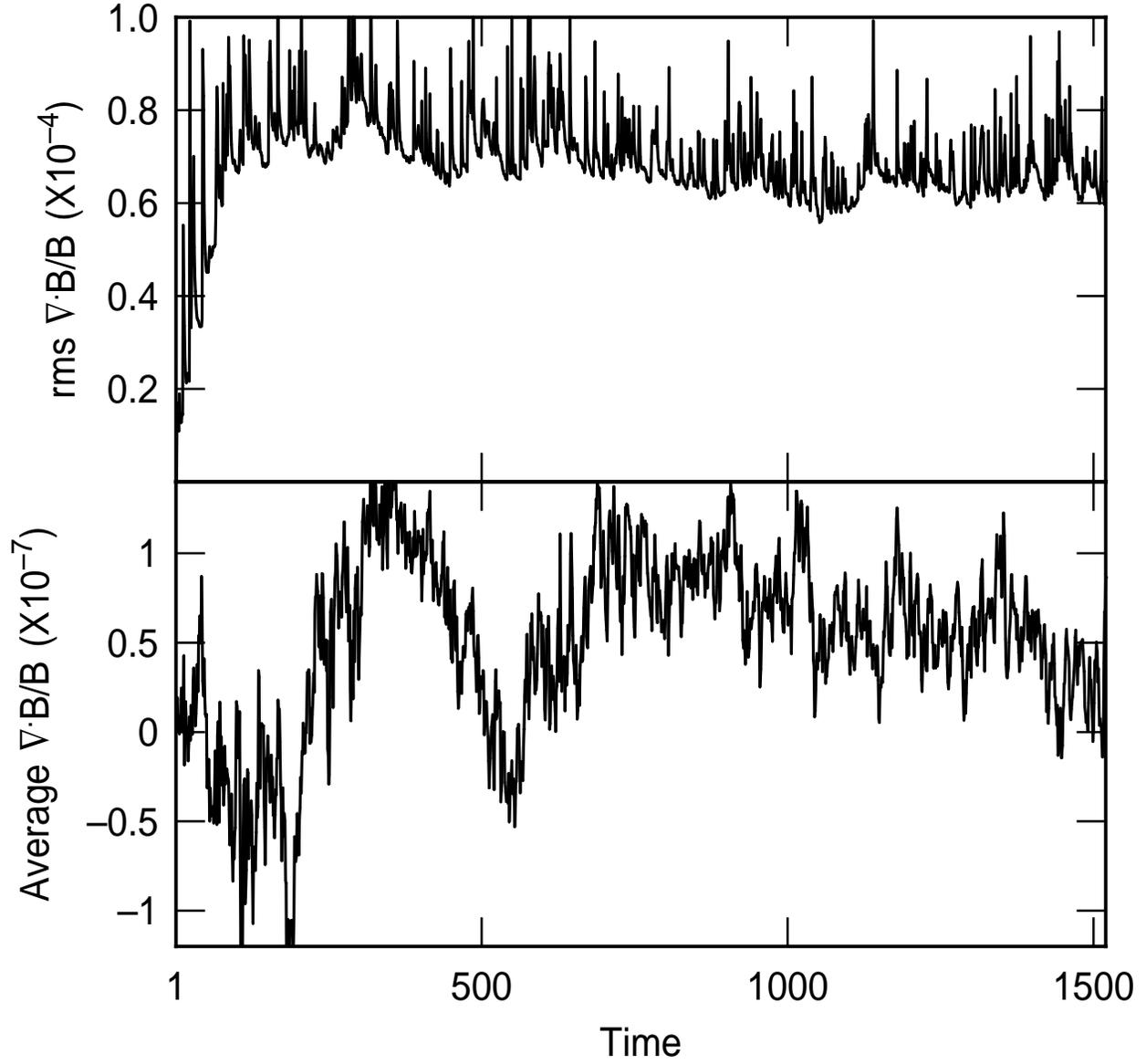}
\caption{The average and rms values of the quantity $\nabla\cdot{\bf
B}/\left({\bf B}\cdot{\bf B}\right)^{1/2}$ are shown as functions of
dimensionless time, with one plotted point per time unit, which is 100
computational time steps.} \label{fig:btest} \end{figure}

Another important test is for the error in $\nabla\cdot B=0$ to be small. Evans \&
Hawley (1988), Stone \& Norman (1992), and Dai \& Woodward (1998)
developed codes that explicitly force $\nabla\cdot {\bf B}=0$, but our
code only gives $\nabla\cdot{\bf B}=0$ to within the numerical accuracy
determined by the grid and time stepping. Figure \ref{fig:btest} shows
the average and rms values of $\nabla\cdot {\bf B}/\left({\bf B}\cdot
{\bf B}\right)^{1/2}$ inside the computational grid of the 2D-MHD
simulation discussed in section \ref{sect:hierarchical}. For this
evaluation, we considered both the total grid, measuring $800\times640$
with the 800 cell direction parallel to the initial field, and the
central portion of this grid, from positions 120 to 680 parallel to the
field (outside the sources of excitation for the waves) and 0 to 640
perpendicular to ${\bf B}$. The values of $\nabla\cdot {\bf
B}/\left({\bf B}\cdot {\bf B}\right)^{1/2}$ are for the latter, shown at
intervals of one time unit, which is 100 time steps, for 1536 time units
overall.

Both the average and the rms values of $\nabla\cdot{\bf B}/\left({\bf
B}\cdot{\bf B}\right)^{1/2}$ are steady throughout the calculation. The
average value is $\pm5\times10^{-8}$ and the rms is
$\sim7\times10^{-5}$. The average is smaller than the rms because this
quantity fluctuates over positive and negative values. There is no
systematic drift for either the average or the rms. 

The lack of any systematic drift in $\nabla\cdot{\bf B}/\left({\bf
B}\cdot{\bf B}\right)^{1/2}$ over time, and the smallness of its value,
imply that stochastic monopoles, which are most likely the result of
round-off errors in calculating {\bf B} and other variables, are
transient and so few in number that they do not affect the code.
Considering that the $560\times640$ grid in which $\nabla\cdot{\bf
B}/\left({\bf B}\cdot{\bf B}\right)^{1/2}$ was measured has
$3.6\times10^5$ cells, there are on average
$\left(7\times10^{-5}\right)\times\left(3.6\times10^5\right)\sim25$
monopoles at any one time, with positive and negative signs canceling
each other to give a net monopole number of less than
$\left(5\times10^{-8}\right)\times\left(3.6\times10^5\right)\sim0.02$.
Even though the code is not designed to force the monopole number to be
zero at all times, this number is still so small that any effects from
non-zero $\nabla\cdot {\bf B}$ are in the noise.

We conclude from this that the magnetic field is sufficiently divergence
free in our simulations to represent the magnetic forces and diffusion
rates with the same accuracy as the other forces. We do not expect the
code to follow all the magnetic field lines to high precision, however,
because of the occasional magnetic monopole.

\subsubsection{Advection test}

\begin{figure} 
\vspace{6.in}
\includegraphics{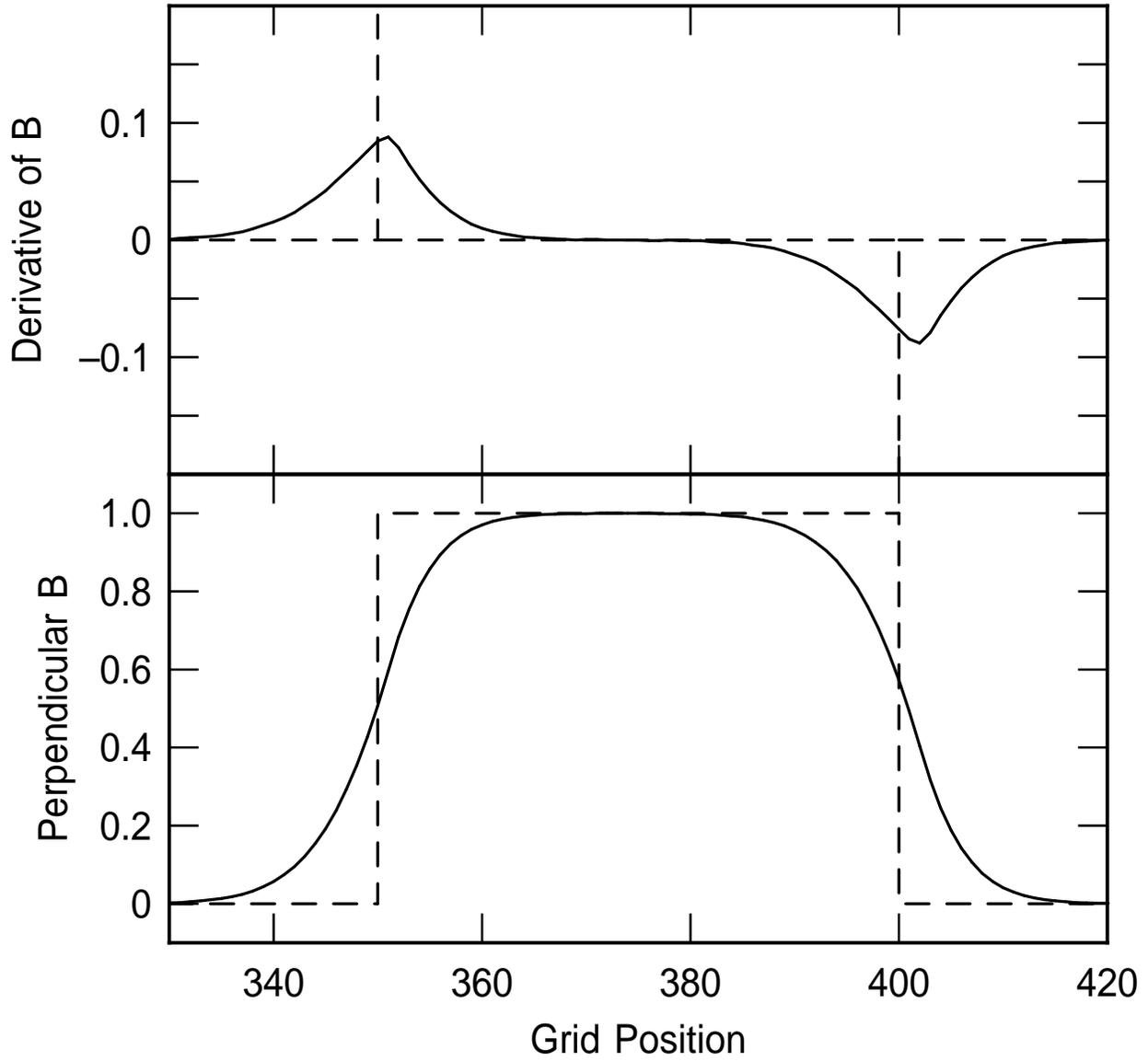}
\caption{(bottom) The perpendicular magnetic field strength versus the grid
position is shown for an initially square magnetic pulse that has been
advected along with a velocity flow field for a distance of five times
its initial thickness. (top) The curl of the field for this pulse, giving
a measure of the sharpness of the edges.} \label{fig:advect} \end{figure}

Other tests for MHD codes were recommended by Stone et al. (1992).
Figure \ref{fig:advect} shows the results of an advection test, in which
the initial conditions are: $B_x=0$, $B_y=1$ between grid points 100 and
150, inclusive, and $v_x=1$, $v_y=0$, $\rho=1$ and $P/\rho=10^{-10}$
everywhere. In this test, a perpendicular magnetic field bundle is
advected through the grid on a current moving at supersonic speed
$v_x=1$. The test is to see how well the initially square pulse
reproduces itself after moving for five times its initial width. A
measure of squareness is the height of the curl $\nabla\times{\bf
B}$, which is $\left(B_y[i+1]-B_y[i-1]\right)/2$ for grid point $i$.
Larger values of $\nabla\times{\bf B}$ at the ends of the pulse
correspond to better advection properties. 

In Evans \& Hawley (1988), four different numerical schemes were tested,
and they gave maximum values of $\nabla\times {\bf B}$ equal to 
0.15, 0.18, 0.07, and 0.03. In Stone et al. (1992), three codes were
tested, and they gave values of 9, 37, and 60 in their figure 2. Stone
et al. used a grid spacing of 0.004 times that used by Evans \& Hawley
and that used here, so we multiply their $\nabla\times {\bf B}$ values
by this factor to get the same scale. The result is then 0.036, 0.15,
and 0.24 for the Stone et al. trials. Here we get a value of 0.088 for
the peak in $\nabla\times{\bf B}$, which is intermediate between the
other tests. The best results were for the piece-wise parallel algorithm
considered by Stone et al.. In the present code, the
spatial derivatives are accurate only to linear order (cf.
\ref{sect:spatial}), so this accounts for the lower $\nabla\times{\bf B}$.

\subsubsection{MHD Riemann test}  

Another test recommended by Stone et al. (1992) is an MHD Riemann
problem with initial conditions: $P=1$, $\rho=1$, and $B_y=1$ for grid
position less than or equal to 400, and $P=0.1$, $\rho=0.125$, and
$B_y=-1$ for positions 401 or larger, with $B_x=0.75$, $v_x=v_y=0$
everywhere. Stone et al. actually had a cell size of 0.125 and a total
number of cells equal to 800, with the divider at the 400th cell, which
is position 50 in his figure 4. Our cell size is 1, so we take 800 total
cells with the divider at 400 in the figure. Stone et al. plotted the
physical quantities at the time $t=10$, which, with our grid,
corresponds to $t=80$ (since the speeds like $P/\rho$ are the same, but
our grid is larger by a factor of 8). Figure \ref{fig:riemann} shows the
results. Each variable is in excellent agreement with the results in
figure 4 of Stone et al. The numbers in the plot of density correspond
to features in the solution (cf. Stone et al. 1992): (1) a fast
rarefaction wave, (2) a slow compound wave, (3) a contact discontinuity,
(4) a slow shock, and (5) another fast rarefaction wave.

Other tests of the basic relaxation method were shown by Jin \& Zin
(1995), without magnetic forces, and without the heating and cooling
functions appropriate for astrophysical problems.

\section{Hierarchical structure in a boundary-free turbulent region}
\label{sect:hierarchical}
\subsection{Excitation of waves}

To experiment with the origin of hierarchical structure in interstellar
clouds, we ran many simulations with moderately strong Alfv\'en waves
generated at the top and bottom of a large grid ($N_y=800$ cells in the
vertical direction plotted here, parallel to the initially uniform B,
and $N_x=640$ cells in the horizontal direction). These waves traveled
towards the center of the grid and interacted there, making an enhanced,
irregular density structure. The boundary conditions were periodic in
both vertical and horizontal directions. Thus the outgoing waves
generated near the top and bottom grid edges meet and mix quickly after
they cross these edges, and the inward moving waves meet after a longer
time in the center of the grid. 

The initial conditions of the grid are a uniform density ($\rho=1$) and
a uniform magnetic field strength ($B_y=\left(4\pi\right)^{1/2}\times
10$; $B_x=0$), giving an Alfv\'en speed of $v_A=10$, a uniform pressure
$P=1$, giving an initial sound speed $\gamma P/\rho=\gamma=5/3$, and zero
velocity in both directions. The gas is also thermally stable initially
($\Gamma=\Lambda$) and the initial cooling time is 1500 time units
($\Gamma_0=10^{-3}$). The magnetic diffusion rate was taken to be
negligibly small, using $\omega_+=N_yv_A/dy$, so the diffusion time
would be $L/v_A$ for a very sharp field perturbation on the scale of the
grid spacing, $dy$, where $L=N_ydy$ is the total grid distance along the field.
Simulations with more rapid diffusion are considered in section \ref{sect:mag}.

To generate waves, the velocities perpendicular to the field, $v_x$, at
certain grid points are changed with a pattern of accelerations
$\partial v_x/\partial t = Ae^{-\omega t}$ for random amplitudes $A$ and
fixed decay rates $\omega$. The spatial positions of these accelerations
are limited to within 5\% and 15\% of the grid size in from the ends
of the grid in the direction
parallel to the initial field. For a grid with $N_y=800$ cells
parallel to {\bf B}, the transverse accelerations occur between cells 40
and 120, and 680 and 760 in the $y$ direction, and throughout
the full width ($N_x=640$ cells) in the perpendicular direction. The
amplitudes $A$ are taken to equal random numbers in an interval from 0
to 1 times some fixed value, chosen to give sufficiently strong
perturbations to make the desired density structures, but not so strong that the
code diverges by forcing a velocity to exceed the previously assigned
maximum velocity $a$ (cf. Sect. \ref{sect:intro}). The decay time is
taken to be $\left(\omega\right)^{-1}=
N_y/\left(4v_A\right)=20$ time units, which is the time it
takes an Alfv\'en wave to move over $1/4$ of the cells parallel to {\bf
B}.

New accelerations are applied continuously, and the old ones terminated
at the same time, so that there is always one acceleration at the bottom
of the grid and another at the top of the grid. The interval between
accelerations is given by a random number between 0 and 1 multiplied by
the time $\left(8\omega\right)^{-1}$. The accelerations at any one time
are confined to a single grid spacing in $y$, parallel to the field,
although different accelerations can occur at different times within the
intervals $(0.05-0.15)N_y$ and $(0.85-0.95)N_y$, discussed above. The
accelerations are also confined to a range of grid points, $\Delta x$,
perpendicular to the field (although each one may extend
beyond the horizontal edges and wrap around to the other side with the
periodic boundary conditions discussed above). 
The range $\Delta x$ has a distribution of
sizes comparable to the distribution of interstellar cloud radii, which
is a power law $n(R)dR\propto R^{-3.3}dR$ (Elmegreen \& Falgarone 1996).
For the simulations, $\Delta x$ varies randomly with this power law
between $0.05N_x$ and $0.25N_x$.

The perturbations are designed to simulate the movements of distant
clouds or other perturbations perpendicular to the magnetic field lines.
These clouds presumably force whole magnetic flux tubes
to move sideways in the manner simulated, since the parallel motions
of clouds do not perturb the field. The cloud motions are also likely to be
supersonic although generally sub-Alfv\'enic in
space, in which case they generate strong enough waves to push matter around and
influence the density, as in the simulation. As long as the waves are sub-Alfv\'enic, they
travel relatively far from their sources (Zweibel \& Josafatsson 1983)
and interact to generate density structures wherever they meet. Some
diffuse clouds, and much of the structure inside both diffuse clouds and the
weakly self-gravitating parts of molecular clouds, can be made in
this way, with remote sources of turbulent energy entering the region
and moving around on magnetic field lines. 

\subsection{Results}

Nonlinear Alfv\'en waves interact to form an intricate density
structure. In the simulations, this structure has many scales because
the initial waves have a power-law width distribution, and because, in
general, non-linear terms in the MHD equations add the spatial
frequencies of two mixing waves, giving an ever-increasing range of
spatial frequencies. 

\subsubsection{Density and temperature maps}

Figure \ref{fig:hier}(top) shows the density distribution of the
simulation described above at a time of $t=1024$ time units. The display
measures $640\times480$, with the 640-cell direction perpendicular to
the initial field and representing the full grid size, and the 480-cell
direction parallel to the initial field and showing only the central
half of the full grid. The density increases monotonically as the color
cycles from blue to yellow to red with the full rainbow, and then jumps
back to blue again, followed by another cycle to red. This cyclical
color display is used to emphasize hierarchical structure. There is also
a density threshold ($\rho<1.21$) below which the figure shows black.
The first blue is at this $\rho=1.21$ threshold, the second is at
$\rho=1.6$, and the highest density is red with
a tiny black dot at $\rho=1.826$, slightly to the right of the middle.
Outside the plotted region, further to the top and bottom, the density
gets moderately low in the ``intercloud'' medium (cells 120--160, and
641--680) and then very low ($\rho=0.19$) in the excitation region,
(cells 40--120 and 680--760). The grid range for the figure is from 161
to 640 cells in the vertical direction.

The density structure in figure \ref{fig:hier}(top) is hierarchical,
with clumps inside larger clumps. The ``cloud'' on the left has three
levels in the hierarchy, making four total levels if the cloud itself is
counted. The density contrast inside the cloud is not large because of
numerical limitations and because of the limited range between the sonic
and Alfv\'en speeds, and between the two equilibrium sonic speeds (both
only factors of 10). The total density contrast in the whole grid is
often a factor of 20 or more, but much of this occurs at the edge of the
cloud where the warm intercloud medium is generated. Higher cloud and
total contrasts ($\sim50$) were achieved in our 1D simulations (Paper
I), because the grid contained 8000 cells and the waves could be driven
harder. Other 2D runs with stronger wave driving made greater density
contrasts, but eventually bombed when a wave velocity in the low density
intercloud medium exceeded the maximum allowed by the parameter $a$ (cf.
Sect. \ref{sect:intro}). Larger values of $a$ degraded the accuracy of
the simulation given the grid size. Future simulations with larger grids
should be able to get around these limitations by allowing larger ratios
between the thermal speeds in the two phases and between the Alfv\'en
speed and the cool thermal speed.

The density structures in figure \ref{fig:hier} are more
ellipsoidal than filamentary, and there are no obvious sharp fronts at
the leading edges of the clumps. This lack of shock-like structures
occurs because, even though the clumps are all moving in bulk, their
speeds are less than the Alfv\'en speed, and also because Alfv\'en waves
are moving through the clumps in both directions, smoothing out the
sharp edges. This gives the model clumps some resemblance to real
interstellar cloud clumps, but this resemblance may only be superficial
because the real structures of interstellar clouds on these near-thermal
scales are not generally resolved.

The complete time evolution of the density structure for this model is
in Figure \ref{fig:mpeg}, which is an mpeg file available from the
electronic version of this paper in the {\it Astrophysical Journal}. The
mpeg file has 192 frames of size $640\times480$. It is a color
representation of the density evolution over a total time of 1529.2
units. Since the initial Alfv\'en crossing time over the entire grid
(800 cells) is $800/10=80$, the simulation represents $\sim19$ Alfv\'en
crossing times. The thermal crossing time in the cool phase, where the
sound speed is $\gamma$, is $\sim800/\gamma\sim480$, so the total time
is 3.2 thermal crossing times (twice this if we consider only the
central cloudy part, which is where the cool gas is). The color code in
the mpeg file is cyclical as in figure \ref{fig:hier}, but there are
three color cycles in density, with blue at $\rho=0.3,$ 1.7, and 2.0.
The maximum density is 2.3, which is red.

The temperature structure at the same time step as in the top of figure
\ref{fig:hier} is shown in the bottom of figure \ref{fig:hier}. 
The highest temperatures at the top and bottom of the grid are black
(threshold $P/\rho>1.7$), and then the temperature decreases
as the colors cycle from red to blue and then again from red to blue. 
The first red is this $P/\rho=1.7$ level, and the second red is at
$P/\rho=1.25$. The minimum temperature, which is blue, corresponds to
$P/\rho=0.9$. 

The intercloud medium at the top and the bottom of the grid is at a much
higher temperature than the cloud in the middle of the grid because the
density is low in the intercloud region, and then the $\Gamma-\Lambda$
function finds the high-temperature equilibrium solution. The clumps
inside the cloud are slightly warmer than the interclump medium because
the clumps are moving compression fronts, not stagnant clouds. This
means the clumps have a continuous source of energy from their
compression. The lower density and temperature in the interclump medium
indicates that {\it the clumps are not confined by an interclump thermal
pressure}, as is often proposed for interstellar clouds. Instead, the
clumps are confined on their leading edges by the ram pressure from
supersonic motions through the interclump medium, and they are confined
on their trailing edges by the gradient of the magnetic wave energy
density that is pushing them along.

The mpeg file shows that the time evolution of the density structure is
similar to what we found for 1D wave compression in Paper I: the waves
push material along with them as they converge in the center of the
grid, and this material builds up to make a ``cloud.'' The cloud has a
low temperature because of its high density, given the heating and
cooling functions, which have two stable thermal states. At the same
time, the waves clear out the matter from the region around the cloud,
leaving an intercloud medium with a low density and a high temperature
equilibrium. The pressure in the cloud is higher than the initial
pressure of the simulation because the incident waves push on the
material they collect in the center. Nevertheless, there is total
pressure equilibrium between the cloud and the intercloud medium, with
the balance between kinetic, magnetic and thermal pressures changing
across the cloud boundary (cf. Paper I). The 1D simulation also showed
how the cloud is broken up into many smaller clouds and clumps, which
have a hierarchical structure. The resolution in the 1D run of Paper I
was 8000 cells, 10 times better than what we have here, so the hierarchy
in density could be seen better there.

The present simulations in 2D show the same cloud formation properties
and hierarchical structure as the 1D runs in Paper I. This structure
changes with time as new waves enter the cloud and the existing waves
continue to interact, but it always has the same hierarchical character.
Three dimensional models will be necessary to fully simulate
interstellar clouds, and it may be that the compression is less for a
given wave amplitude in 3D than in 2D, because of the additional degree
of freedom for magnetic motions in 3D. Larger compressions can always be
applied to get the same level of density enhancements. The formation of
hierarchical structure should not depend on the dimensionality of the
simulation, however. It seems to be characteristic of non-linear wave
interactions in any number of dimensions.

The mpeg file indicates that the small clumps exist for a shorter time
than the large clumps. The lifetimes of clumps of various sizes are
estimated to be about the sound crossing time inside the clump, regardless of
scale. This lifetime is
definitely larger than the internal Alfv\'en crossing time. This
result is consistent with the view presented below in section \ref{sect:mag}
that most of the clumps are sonic or mildly supersonic features created
by thermal pressure gradients parallel to the field.  

\subsubsection{Power Spectra}

To quantify this hierarchical structure, we measured the Fourier
transform of the density in directions parallel and perpendicular to the
initial magnetic field over lengths of 240 cells and 640 cells,
respectively (the FFT double-precision subroutine from the IBM ESSL
Fortran library accommodates vectors with 640 or 240 elements). 
The vertical length of 240 cells was chosen to represent the
inner part of the cloud; this is the central half of the vertical
extent of the grid in figure \ref{fig:hier}.

For the FFT in the horizontal direction, perpendicular to the initial
field, separate FFTs of the horizontal density distributions were made
for each vertical grid position between cells 280 and 520, and averaged
together. For the FFT in the vertical direction, separate FFTs of the
vertical density distributions were made for each horizontal grid
position between 0 and 640. 

The results for the timestep shown in figure \ref{fig:hier} are in
figure \ref{fig:fft}, which plots the Fourier transform power,
$\left(Re^2+Im^2\right)^{1/2}$ for real and imaginary parts, Re and Im,
versus the spatial frequency. The FFTs for the directions parallel and
perpendicular to the field are shown on the bottom and top,
respectively. The spatial frequency on the bottom figure equals 240
cells divided by the wavelength of the Fourier component, measured in
units of the grid spacing. The spatial frequency on the top equals 640
divided by the wavelength. Each plot goes from a spatial frequency of 1,
which is for a wave spanning the whole extent of the corresponding
direction, to 120 or 320, respectively,
which are both for a wavelength of 2 cells.

The sources of excitation contribute somewhat to the FFT in the
perpendicular direction. These sources have a power law size
distribution between lengths of 32 and 160, which correspond to spatial
frequencies of 20 and 4 in figure \ref{fig:fft}. The top diagram in that
figure has a slightly shallower slope in this range than at higher
frequencies. Below a spatial frequency of 20 in the perpendicular
direction, the slope is $\sim2.5$; above 20 it is $\sim3.6$, and in the
parallel direction, between 10 and 100, it is $\sim3.2$. 

For comparison with these model results, Green (1993), 
Stutzki et al. (1998), and Stanimirovic et al. (1999)
found a slope of $-2.8$ for
the power spectrum of the line-of-sight integrated intensity structure
in HI and CO maps. The 2D model results are not expected to be the same
as this, but the fact that both real clouds and the model results give
fairly smooth power-law power spectra indicates that both have
self-similar structure on a range of scales spanning at least a factor
of 10.

\subsubsection{Velocity Correlations}

The rms velocity in a region of our simulation increases with the size
of the region, as for Kolmogorov turbulence. Figure \ref{fig:vel} shows
the average rms velocities parallel and perpendicular to the initial
magnetic field in squares of various sizes, as indicated by the
abscissa. The time step is the same as in figure \ref{fig:hier}. The rms
velocity scales as a power law with the size, $S$, $v_{rms}\propto
S^{\alpha}$, with $\alpha\sim1$ on scales smaller than $\sim30$ cells,
and $\alpha\sim0.3$ on scales from 30 to 240 cells. The largest box size
considered is 240, which is the vertical extent of the cloudy part of
the simulation; this is much smaller than the size of the grid
($800\times640$), so edge effects are not likely to influence this
velocity correlation. Evidently, the largest scales have a velocity-size
correlation similar to Kolmogorov turbulence, with a slope of about
$1/3$. 

Smaller scales have a steeper correlation slope because the velocity
differences tend to be too small on small scales. This means that the
material tends to move too coherently compared to Kolmogorov turbulence
on scales less than about 30 cells. The origin of this steepening could
be numerical: although 30 cells should be sufficiently large to be free
of resolution errors at the cellular scale, the resolution of shocks and
rarefaction fronts is $\sim5-20$ cells in figure \ref{fig:riemann},
depending on shock strength. Also, the range of scales for driving the
waves is 32 to 160, which is in the flat part of the velocity
correlation. 

A physical origin for the steepening at small scales is likely too,
because small regions are forced to co-move by the magnetic field
(Parker 1992). Indeed, the Alfv\'en speed is $\sim10$ in the turbulence
simulation, and this is much larger ($\times100$) than the rms speed
where the velocity-size correlation becomes steep in figure
\ref{fig:vel}. This implies that magnetic field tension may be
sufficiently strong to overpower the inertial forces from turbulence on
small scales. If this is also the case in interstellar turbulence (and
the 2D model results still apply in 3D), then the velocity-size
correlation in molecular clouds should become steeper than the usual
$\sim0.4$ power law slope at linewidths less than some small fraction
($\sim$1\% in these simulations) of the Alfv\'en speed. To observe this,
the total range in clump rms turbulent speeds must exceed a factor equal
to the inverse of this fraction, considering that the broadest linewidth
is usually the Alfv\'en speed.

Figure \ref{fig:vel} also indicates that the rms velocities parallel
to the magnetic field are smaller than they are perpendicular to the
field. This is because the motion is forced in the perpendicular 
direction, and the parallel motion responds as a higher order (non-linear) 
effect.  It may be that these two speeds are more similar in real
interstellar clouds because all of the motions there are expected
to be more non-linear.  There could be some decoupling between the
compressive and shear waves too (Ghosh et al. 1998).

\subsubsection{Summary}

Interacting non-linear magnetic waves can make hierarchical, scale-free density
structure with an approximately Kolmogorov velocity-size relation out of
an initially uniform medium. The clouds and clumps that are produced by
this mechanism are similar to what is observed in diffuse and
translucent interstellar clouds, and in the parts of molecular clouds
that are not strongly self-gravitating. Interstellar cloud structure is
therefore likely to be partly the result of non-linear magnetic wave
interactions, which is sub-Alfv\'enic MHD turbulence. 

\section{Experiments with Magnetic Damping}
\label{sect:mag}

There have been several suggestions that stars begin to form when the
minimum length for magnetic waves in the presence of ion-neutral
diffusion becomes larger than a Jeans length (e.g., Mouschovias 1991).
This idea is based on a model in which stars form in a more-or-less
uniform cloud that is supported against self-gravity on large scales by
MHD turbulence. When MHD turbulence is no longer possible, which indeed
happens on sufficiently small scales in a uniform cloud, gravity wins
and the local region collapses. 

Our model of star formation is very different, because it proposes that
clouds are never uniform. Turbulence from either inside or outside the
cloud always gives them high-contrast density structure over a wide
range of scales. Star formation begins when this structure melts away on
scales larger than the thermal Jeans length. When there is such
pervasive structure, particularly with the molecular cloud correlations
between density, velocity dispersion, and size, magnetic diffusion does
not become relatively more important on small scales. In fact the ratio
of the diffusion time to the wave time is about constant on all scales
for such a model. This is because smaller regions are denser, and so the
neutral gas is more tightly bound to the ions during field line motions
in just the right amount to compensate for the heightened magnetic
tension (Elmegreen \& Fiebig 1993). A cutoff at small scales finally
arises when the density is so high that the small grains stop gyrating
around the field.

The present code cannot check these ideas directly because there is
no self-gravity. Instead, we assessed the importance of magnetic
diffusion on cloud structure in a different way, with a 
series of experiments having different
amounts of magnetic diffusion, adjusted through the parameter
$\omega_+$; this is the ion-neutral collision rate introduced in
equation (\ref{eq:vdiff}). Two runs are compared here. Both models had
exactly the same parameters and random numbers in a $400\times320$ grid,
and they had the same solutions up to the time 528.2 time units. This
time equals 13.2 Alfv\'en wave crossing times through the vertical grid,
and 1.32 sound crossing times for the cool phase, which is long enough
to get a cloud in the center. Both runs also had
$\omega_+=10^{-2}N_yv_A/dy=40$ up to this time, but then one of them
continued after this time with the same $\omega_+$ and the other
continued with $\omega_+=10^{-3}N_yv_A/dy=4$. These diffusion rate
constants correspond to diffusion times of
$\omega_+/\left(kv_A\right)^2$ for field gradient $kB$, and this time
equals $\omega_0\left(L/v_A\right)\left(kdy\right)^{-2}$, where
$L=N_ydy$ is the full length of the grid along the field, $dy\equiv1$ is
the grid spacing, and $\omega_0=10^{-2}$ and $10^{-3}$ in the two cases,
respectively. This means that the diffusion time is $\omega_0$ times the
product of initial Alfv\'en wave crossing time and the square of the
scale length for magnetic field gradients, measured in grid spacings.
The simulation discussed in section \ref{sect:hierarchical} had a large
$\omega_+=N_yv_A/dy$, which gave a diffusion time equal to the Alfv\'en
crossing time times the square of the scale length. 

The case with rapid diffusion, $\omega_0=10^{-3}$, noticeably lost the
perpendicular component of the field inside the cloud, which means that
the wave amplitude dropped even though an external excitation with the
same amplitude was still applied. The other run, with
$\omega_0=10^{-2}$, continued with a high internal wave amplitude,
following the same excitation from outside the cloud. To follow the
rapid diffusion in the first case, we had to decrease the time step by a
factor of 10 following this transition at $t=528.2$; to be consistent in 
the second run, the timestep was decreased there too. After
the transition, there were 202000 more time steps for each run, or an additional
simulation time of 202 time units, which is 5 initial Alfv\'en wave
crossing times through the whole grid.

The resulting density maps (not shown) had surprisingly little
differences in the two cases -- they were virtually indistinguishable,
even though the wave amplitude in the case with rapid diffusion got to
be 5 times less than in the other case. 

We demonstrate this result in two ways. Figure \ref{fig:rmsbrho} shows
the time development of the rms density and the rms of the perpendicular
component of the field, with the rapid diffusion case represented by
dashed lines. These rms values are taken along the horizontal rows in
the grid, perpendicular to the field. This avoids the overall gradient
in the vertical direction from the general cloud structure. Thus, the rms values of
the density and perpendicular field component were determined for each of
the 121 horizontal rows in the middle of the grid, with each rms
calculated from all 320 grid values in the horizontal rows.
All these rms values
were then averaged over the 121 rows to get a single rms value at each interval of 1.6 time 
units (every 1600 time steps). The results are plotted in figure \ref{fig:rmsbrho}.
The figure shows one full cycle of the cloud's overall density oscillation (which is still
a response from the initial pulse of high pressure during cloud formation).
The rms of the density is a measure of the strength of the clumpy
structure; it is nearly the same in the high diffusion case as it is in the
low diffusion case.  The rms of the perpendicular field is a measure of
the wave amplitude inside the cloud. It starts the same in the two cases, 
but gradually dies away in the high diffusion case, ending up about a
factor of 5 weaker than in the low diffusion case.  Thus {\it the internal
magnetic waves die out, but the clumpy structure is unchanged.}

The same result is shown again in figure \ref{fig:fftbrho}, now using
Fourier transform power spectra to measure the strengths of the density
and wave perturbations. These diagrams were made like figure
\ref{fig:fft}, but now for a grid that is smaller in each direction by a
factor 2. The two left-hand diagrams show the average power spectra of
the density in the region of the cloud (120 cells parallel to the field,
out of 400 total, and the full 320 cells perpendicular to the field) in
the parallel-to-field direction on the bottom and perpendicular
direction on the top. The two right-hand diagrams show the power spectra of
the perpendicular component of the magnetic field. There are three lines
in each diagram: the solid line is at the beginning of the experiment,
at the time $t=529.2$ time units, the long-dashed line is at the end of
the experiment with little magnetic diffusion ($\omega_0=10^{-2}$),
i.e., at $t=730.2$ time units, and the short-dashed line is at the end
of the experiment with a high rate of magnetic diffusion
($\omega=10^{-3}$).  Both of the top diagrams also show linear fits
to the power spectra from spatial frequencies of 20 to 160, shifted
upwards by factors of 100 for clarity. These linear fits reveal the
similarities and differences between the power spectra without the
noise.

Evidently, the density power spectra are virtually
indistinguishable at the ends of the two experiments. The power spectrum
of the perpendicular component of the field measured along the field
(lower right diagram) is also nearly the same in the two cases; this
means that the wave structure along the field is about the same with high and low
diffusion.  This result is not surprising because both experiments were driven
with exactly the same incident waves, and both had the same wave
propagation speeds through the grid. However, the power spectrum of the
perpendicular component of the field measured across the field is much
less at high spatial frequency in the high diffusion case than in the
low diffusion case (see the top right diagram where the
short-dashed line is below the others). This means that the wave
amplitude is lower after some time when the diffusion rate is high, as
expected for magnetic waves in general. 

What is perhaps surprising about this experiment is that even though the
magnetic wave amplitude gets low after some time in the high diffusion
case, the density structure is virtually unchanged. This means that
enhanced magnetic diffusion does not lead to a loss of cloud structure,
even when this structure is directly the result of magnetic wave
interactions. How can this be?

The reason for this result is that the density structure in all of the
cases considered in this paper comes from motion along the magnetic
field that is driven by pressure gradients in this direction. Inside the
cloud, this compression is sonic in nature, because of the dominance of
the $\nabla P$ term at high density in the parallel momentum equation.
This is true even when the material is pushed at supersonic (but
sub-Alfv\'enic) speeds. The origin of the motion is the noise at the edge
of the cloud, which is subject to really strong magnetic waves from
outside. The waves themselves weaken and, in the high diffusion case,
damp out, as they travel through the high density part of the cloud, but
their damage has already been done long before this. The primary
influence of the external waves is at the cloud edge, where the magnetic
energy and the momentum of interclump motions get converted into cloud
density pulses, like puffs of wind, 
that travel through and interact with each other in the
interior of the cloud. Even when the internal magnetic wave energy is
damped, these sonic pulses still make essentially the same
density structures inside the cloud. 

If the density structure inside real interstellar clouds is the result
of interacting, non-linear magnetic waves, as in the models discussed
here, then this structure would seem to be relatively unaffected by an
enhancement in magnetic diffusion that might result from an increase in
density or an excess shielding of external radiation. This result
suggests that enhanced magnetic diffusion is not the key to the onset of
star formation in weakly self-gravitating clouds. Of course, magnetic
diffusion can still play a very important role later, during the
accretion phase inside a strongly self-gravitating cloud piece, but this
process is not simulated here. 

\section{Experiments with Gravitational Density Gradients}
\label{sect:grav}

In a model where clouds and clumps form by interacting non-linear
magnetic waves, the only way the tiny structure can disappear as a
necessary precursor to star formation is if both the magnetic waves and
the sonic pulses they create damp out before they reach the cloud
center. The previous section showed that even when the waves damped out,
the sonic pulses that were generated in the intercloud medium and at the
cloud edge still remained. 

Here we consider a different way to damp the sources of internal cloud
structure. This occurs when both the external waves and the sonic pulses
have to climb up a steep density gradient before getting to the center.
The increased density removes the wave and pulse kinetic energy by
momentum conservation, and this leaves the center with relatively little
turbulence to drive structure formation. Such a cloud density gradient
occurs naturally when the cloud becomes significantly self-gravitating.
This means that {\it the gradual contraction of a cloud under the
influence of self-gravity should be enough to exclude externally
generated turbulence and initiate the decay of tiny cloud structure in
the core}. 

This effect is demonstrated in two ways. First WKB solutions to the wave
equation are given for waves traveling through a region with a
centralized density enhancement. These analytical solutions show the
expected decrease in wave amplitude in the cloud center. Second, two
numerical experiments are run on $800\times640$ grids that have a fixed,
plane-parallel gravitational force in the direction along the field,
which gives them a $\rho=\csc^2\left[\left( y-y_0\right)/H\right]$
general density structure underneath the wave structure. These two
experiments have different scale heights, $H$, and when combined with
the simulation discussed in Section \ref{sect:hierarchical}, show a
gradual loss of density structure as the scale height decreases and the
central density increases.

\subsection{WKB Solutions to waves in density gradients}

We consider here a simple wave of any kind that satisfies the wave equation
\begin{equation}
{{\partial^2 W}\over{\partial t^2}}=a^2{{\partial^2 W}\over{\partial y^2}}
\end{equation}
for wave amplitude $W$ and wave speed $a(y)$ that is a function of position, $y$.
Using the WKB approximation for weak waves, we write
\begin{equation}
W(y,t)=e^{i\omega t-i\int_{-\infty}^y k dy}\end{equation}
for frequency $\omega$, wavenumber $k=2\pi/\lambda$ and wavelength $\lambda$.
Substituting this waveform into the wave equation gives
a differential equation for $k(y)$:
\begin{equation}{{\omega^2}\over{a^2}}=ik^\prime+k^2\end{equation}
for derivative $k^\prime=dk/dy$.
Substituting the real and imaginary components for complex $k=k_r+ik_i$ then
gives two equations, one real and the other imaginary. We look for
pure wave solutions with real $\omega$, and this allows us to eliminate
one equation, giving as a result a single equation for the real component
of $k$:
\begin{equation}
k_r^{\prime\prime}-{{3\left(k_r^\prime\right)^2}\over{2k_r}}
+2k_r^3-\left({{\omega^2}\over{a^2}}\right)2k_r=0.
\label{eq:wkb}
\end{equation}
The imaginary component
of $k$ was eliminated from the above equation, but 
is given by $k_i=-k_r^\prime/\left(2k_r\right)$.

Equation \ref{eq:wkb} was solved numerically for $k_r(y)$. The wave
speed is taken to be \begin{equation}
a(y)={{e^{y/H}+e^{-y/H}}\over{e^{y_e/H}+e^{-y_e/H}}}, \end{equation} so
it equals unity at the edge of the numerical grid, where $y\equiv
\pm y_e=5$, and there is a gradual slow down of the wave to a minimum wave
speed of $2/\left(e^{y_e/H}+e^{-y_e/H}\right)$ at $y=0$. In the cloud
model, this slow down is the result of an increased density. The
desired result is the ratio of the wave amplitude at the center to the
incident wave amplitude at the edge. 
The boundary condition for the integration is $k_r=\omega/a$ at
$y=-y_e$.

Figure \ref{fig:wkb} shows the result. The average wave amplitude inside
the central scale height of the grid, between $y=\pm H$, is shown as a
function of the central wave speed. The wave amplitude decreases as the
central wave speed decreases, almost exactly as the square root for this
model; i.e., $<W>\approx a(0)^{1/2}$. 

This decrease in amplitude with increasing density is essentially the
result of gradual wave reflection at the cloud edge. The net flux toward
the cloud on each side is the difference between the incident and 
reflected wave fluxes, and by conservation, must equal the respective
fluxes in the same directions inside the cloud. We checked the WKB
result by considering a cloud/intercloud boundary with a sharp immovable
edge and a hard barrier inside the cloud to simulate reflection symmetry
through the cloud (this is analogous to our MHD solution, which has
waves incident from both sides of the cloud). We calculated the
reflection and transmission amplitudes of the waves, and then averaged
the internal and external wave energy densities over a factor of 100 in
wavenumber (to smooth out resonances). We found again that the average
wave amplitude inside and outside the cloud is always proportional to
the square root of the local Alfv\'en speed.

This result differs from the proposal by Xie (1995) that
Alfv\'en waves maintain an amplitude inversely proportional to the
square root of density. In our case, the total field strength is about
constant and the wave amplitude varies as the inverse fourth root of
density. This implies that the wave pressure and energy density are
lower inside the cloud than outside, demonstrating the effect of
shielding. There is also a net compression of the cloud from this
shielding, rather than a wave-pressure equilibrium inside and outside
the cloud, as there would be in the case considered by Xie. 

\subsection{MHD solutions to waves in density gradients}

The same problem was studied with the MHD code. We ran two more 
2D simulations as in section \ref{sect:hierarchical} but with
an additional acceleration from constant gravity, directed toward
the center of the grid in the vertical direction, along the field. 
The gravitational acceleration, $g$, was written as part of
the equation of motion in equation \ref{eq:motion}.  Here it given by
\begin{equation}
g(y)=-\left({{2a_0^2}\over{H}}\right)\left({{e^{\left(y-y_0\right)/H}-e^{-\left(y-y_0\right)/H}}
\over{e^{\left(y-y_0\right)/H}+e^{-\left(y-y_0\right)/H}}}\right),
\end{equation}
for initial isothermal speed $a_0$ given by $P/\rho=a_0^2=1$.
The corresponding initial condition for density was taken
to be the equilibrium value
\begin{equation}
\rho(y)=\left({{e^{-y_0/H}+e^{y_0/H}}
\over{e^{\left(y-y_0\right)/H}+e^{-\left(y-y_0\right)/H}}}\right)^2.
\end{equation}
The midpoint of the grid in the vertical direction is $y_0=400$.  This
density is normalized to equal 1 at the top and bottom edge of the
grid ($y=2y_0$ and 0, respectively), 
as in section \ref{sect:hierarchical}, but now the density is
higher in the center by a factor that depends on the scale height. 
We chose one run with $H=300$ cells, giving a central density
enhancement of 4.1, and another with $H=235$, giving a
central density of 8.0.

The simulations were run for as long as that in section
\ref{sect:hierarchical}, with the same random numbers and external wave
stimulations. This allowed time for internal adjustments and
large-scale cloud oscillations.  The results show that the
transverse wave velocity inside the cloud, and the level of
density fluctuations, both decrease as the central density from
gravity increases.  Figure \ref{fig:grav} shows the power spectra
of the density at the bottom and the transverse velocity at the top for the
three runs indicated by different line types.  These power spectra
were taken from Fourier transforms in the transverse direction, 
as on the top of figure \ref{fig:fft}.  The density power spectra 
are normalized to the
power at zero spatial frequency to remove the differences between
the absolute densities in the clouds in these cases (recall that the
central density is $\sim2.5$ in the gravity-free case, while it is $\sim4$ and
$\sim8$ in the two gravity cases). The sloping lines in the figures are least
squares fits to the power spectra between frequencies 20 and 320, 
shifted upwards by factors of 100 for clarity.  The fits indicate
more clearly than the noisy power spectra that the gravity-free
case, indicated by the solid line, has more power in both transverse
wave velocity and density structure than the two centrally condensed
cases.  

The downward shift in the velocity power spectrum when gravity is added is a factor
of 2.1 for the case with a central density of 4 and 
3.1 for the case with a central density of 8, measured at a
spatial frequency of 20.  These numbers 
compare well with the results in figure \ref{fig:wkb}, considering that
the minimum wave speed is proportional to the inverse square root of the
density. The downward shifts in the density power spectra at a frequency of 20 are
factors of 3.0 and 3.7, respectively.  The degradation in density structure
is greater than in velocity structure, as expected for density caused by
compression. If the wave energy density completely dominated the thermal pressure,
then the density variations would scale as the square of the velocity
variations. 

The power spectra of velocity and density in the direction parallel to
the field do not change much when the cloud becomes centrally condensed.
This is analogous to what was found for the magnetic diffusion tests,
where only the perpendicular direction had any change in magnetic wave
amplitude. This is the result of a similar wave structure parallel to
the field in all the cases of central condensation, following from the
same wave stimuli. The primary differences between the runs 
are in the wave amplitudes. 

These results indicate that {\it the internal velocity and clumpy structure
begins to disappear as the overall cloud gets more and more centrally
condensed from gravity.}

\section{Conclusions}

The Jin \& Xin (1995) relaxation algorithm has been adapted to do MHD
simulations with an energy equation and magnetic diffusion term relevant
to astrophysical problems. The code tested well in comparison with
other astrophysical MHD codes, and was judged to be adequate for the
problems considered here.

MHD simulations in two dimensions were run for several cases to address
the question of how hierarchical and scale-free clumpy structure inside
interstellar clouds and in the general interstellar medium might be
created, and how this structure might be induced to go away on very
small scales as a precursor to star formation. 

The simulations in section \ref{sect:hierarchical} showed that
interacting non-linear Alfv\'en waves can make a whole 2D cloud and
self-similar clumpy structure inside of it. This structure was
demonstrated by its power-law power spectrum and by its overall
hierarchical appearance. The clumps were rounded rather than
filamentary, with no obvious shock structures, and each one existed for
about one internal sound crossing time, regardless of scale. The thermal
pressure of the interclump medium does not confine the clumps. They owe
their existence entirely to kinematic pressures that are associated with
their motions and with the relative motions of the surrounding gas. The
overall structure had a realistic Kolmogorov velocity-separation
relation at large scales, but it had a steeper relation at small scales
for unknown reasons.

The simulations in section \ref{sect:mag} showed that enhanced magnetic
diffusion caused the Alfv\'en wave amplitude to decrease inside the
cloud, but it did not change either the amplitude or the pattern of the
density structure. This experiment suggests that, on interstellar
scales which span
the range between subthermal and superthermal motions, the density structure
comes primarily from strong sonic pulses that move along the field and
interact in complex ways. Changes in the magnetic diffusion rate do not
change these density structures as long as the sonic pulses that drive
them continue to have a source. 

The simulations in section \ref{sect:grav} showed that both the internal
waves and the small-scale density structures they help create go away
when the overall cloud has a density gradient, as might result from bulk
self-gravity. This is because externally generated waves have trouble
penetrating a cloud and making small scale structure by non-linear
interactions when they have to climb a strong density gradient first. A
WKB solution to the wave equation obtained the same result.

We view the processes acting in these simulations to be at the bottom
end of the range of scales of interstellar cloud structures, near the
thermal pressure limit where turbulent pressures are only slightly
larger than thermal. We believe that a larger computational grid would
show a more extended hierarchical structure on larger scales, but the
same wave interaction processes on small scales. Most observations of
real interstellar clouds cannot yet resolve the small scales that are
simulated here, but only the larger parts of the hierarchy of cloud
structure. This implies that if the models are a guide to reality, then
the supersonic turbulence that is observed in interstellar clouds is not
really supersonic at the atomic level, but nearly thermal locally, i.e.,
with relatively small local velocity gradients and few strong shocks. In
that case, the appearance of supersonic motions is the result of a
superposition of locally near-thermal motions on a wide range of unresolved
scales, with a Kolmogorov-type velocity-size correlation generating the
largest speeds.

The simulations also suggest a mechanism for star formation that is
relevant to the modern model of interstellar clouds, in which many
clouds and the clumpy structures inside them come from magnetic
turbulence and the associated non-linear wave interactions, and in which
these structures initially
extend down to very small scales, far past the scale of
the minimum stellar mass. Star formation in such a cloud model does not
require any fragmentation mechanism, nor any other mechanism that
separates out stellar mass units from the background, because these
clouds are always highly fragmented anyway, on all scales, including the
range covered by stars. That is, stellar mass fragments are in all
clouds all the time, as a result of turbulence. Star formation with
such a cloud model requires a smoothing mechanism instead, one in which
sub-stellar fragments coalesce and meld together to build up smooth
pools with stellar masses, without the destructive and dividing influence
of turbulence inside and around this pool. 

The simulations suggest that enhanced magnetic diffusion is probably not
important for this first step in star formation, but the formation of a
gradual density gradient from bulk self-gravity is. This result leads to
the following scenario for star formation:

Star formation begins in a region of interstellar gas when various
processes render it so massive that gravitationally driven
motions become comparable to the externally and internally driven
turbulent motions. At this time, the cloud begins to contract under
its own weight and builds up a density gradient. Such a density gradient
will shield the cloud from turbulence in the outside world, and lead to
a reduction in the internal turbulent energy, as well as a loss of
internal small-scale structure. Because the smallest scales evolve the
quickest, this loss of structure will begin on the smallest scales and
quickly increase the mass of the smallest smooth, thermally-dominated
clumps. When these smallest, thermally-dominated masses increase to the
point where gravitationally driven motions inside of them begin to
dominate thermal motions, they collapse catastrophically to make one
or a few stars each in dense cores. Neighboring clumps do the same,
all rather quickly on the scale of the overall cloud evolution, forming
a hierarchical arrangement of stars and star clusters on time scales
comparable to the turbulent crossing times for those
scales. If the self-gravity of the overall cloud is very strong, then
this resulting cluster will be very dense, if not, then only a sparse
cluster will form. This is in agreement with the densities, structures and
formation times of real star clusters (see review in Elmegreen et al. 1999).

At the present time, these initial star-formation processes are beyond
the limit of angular resolution in general cloud surveys. They also
should occur quickly on the first substellar scales, making the initial
smoothing process unlikely to see. This implies that interstellar clouds
with and without star formation may look very similar on today's
observable scales, and their difference only show up when the angular
resolution is great enough to see structures 
at $\sim500$ pc with far less than a
stellar mass. Our prediction is that star-forming clouds will have much
less structure on sub-stellar scales than pre- or non-star forming
clouds, and that extremely young star-forming regions will have a
relative number of substellar clumps that is midway between those of the
non- and the active star-forming clouds. That is, the clump mass
spectrum will change from a power law to very small scales in
non-star-forming clouds to one with a flattened or turned-over
distribution function at low mass in pre- or active star-forming clouds,
and the mass at this flattening or turnover will increase up to the
minimum stellar mass as the cloud becomes more and more active with star
formation. Another signature of this process is the hierarchical
structure of young stellar positions, whatever the scale and density of
star formation, and the appearance of clusters rather quickly, in about
a local crossing time. 

Acknowledgements: Helpful comments on the manuscript by Dr. A. Lazarian
and S. Shore are appreciated.

\newpage

\newpage

\newpage

\newpage
\begin{figure} 
\vspace{6.4in}
\includegraphics{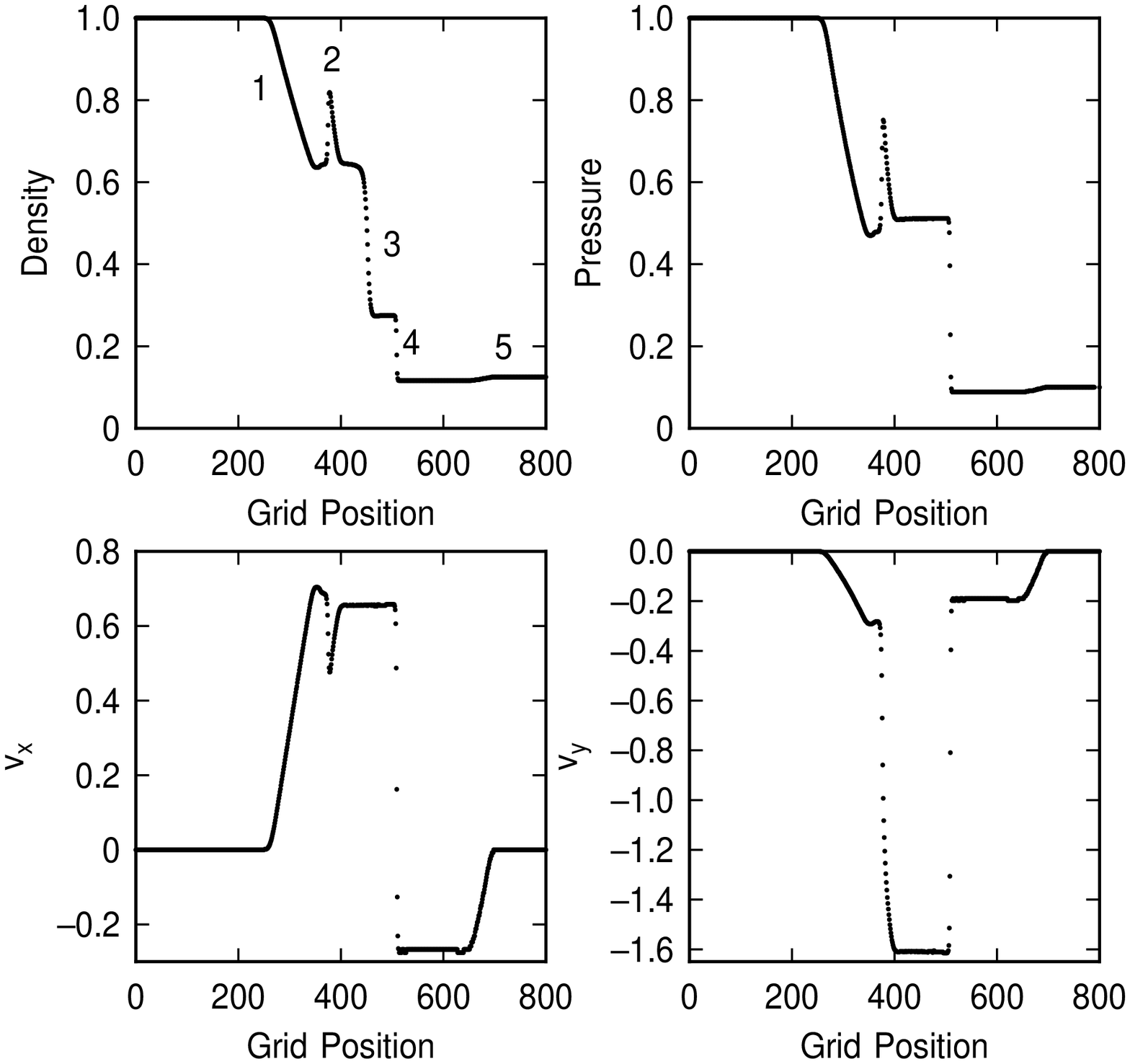}
\includegraphics{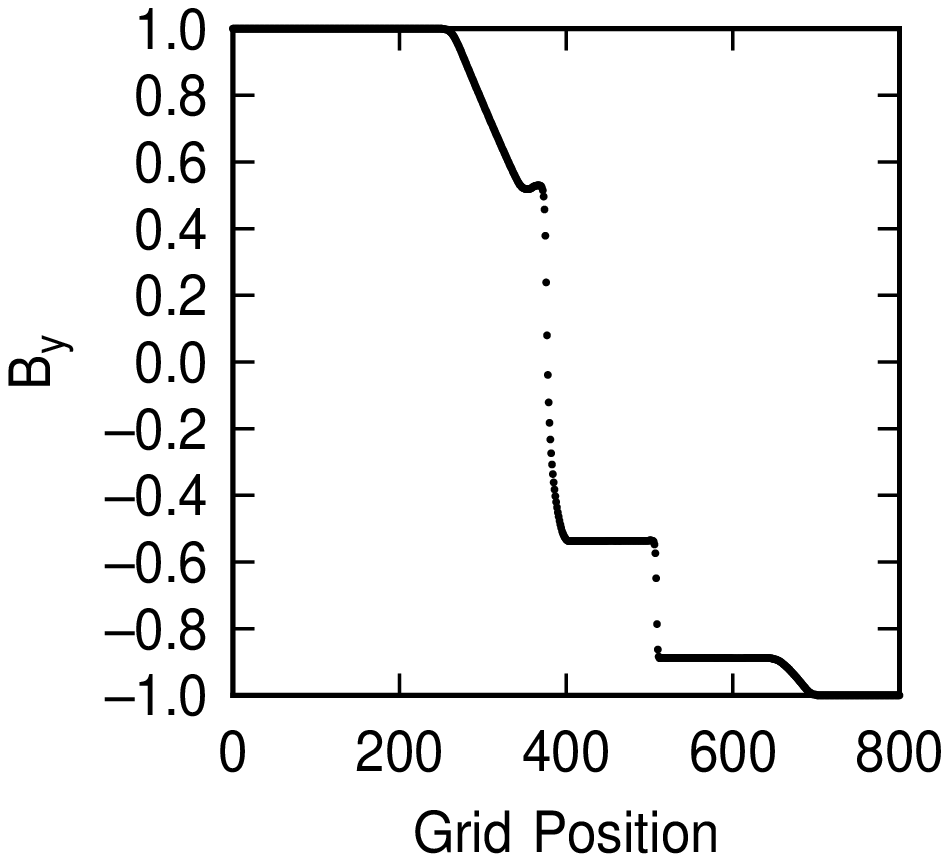}
\caption{Physical variables in the MHD Riemann test discussed by
Stone et al. (1992). The test here matches well the results in Stone et
al.. The five main features in these figures are (1) a fast rarefaction
wave (2) a slow compound wave, (3) a contact discontinuity, (4) a slow
shock, and (5) another fast rarefaction wave.}
\label{fig:riemann}
\end{figure}

\newpage
\begin{figure}
\vspace{6.3in}
\includegraphics{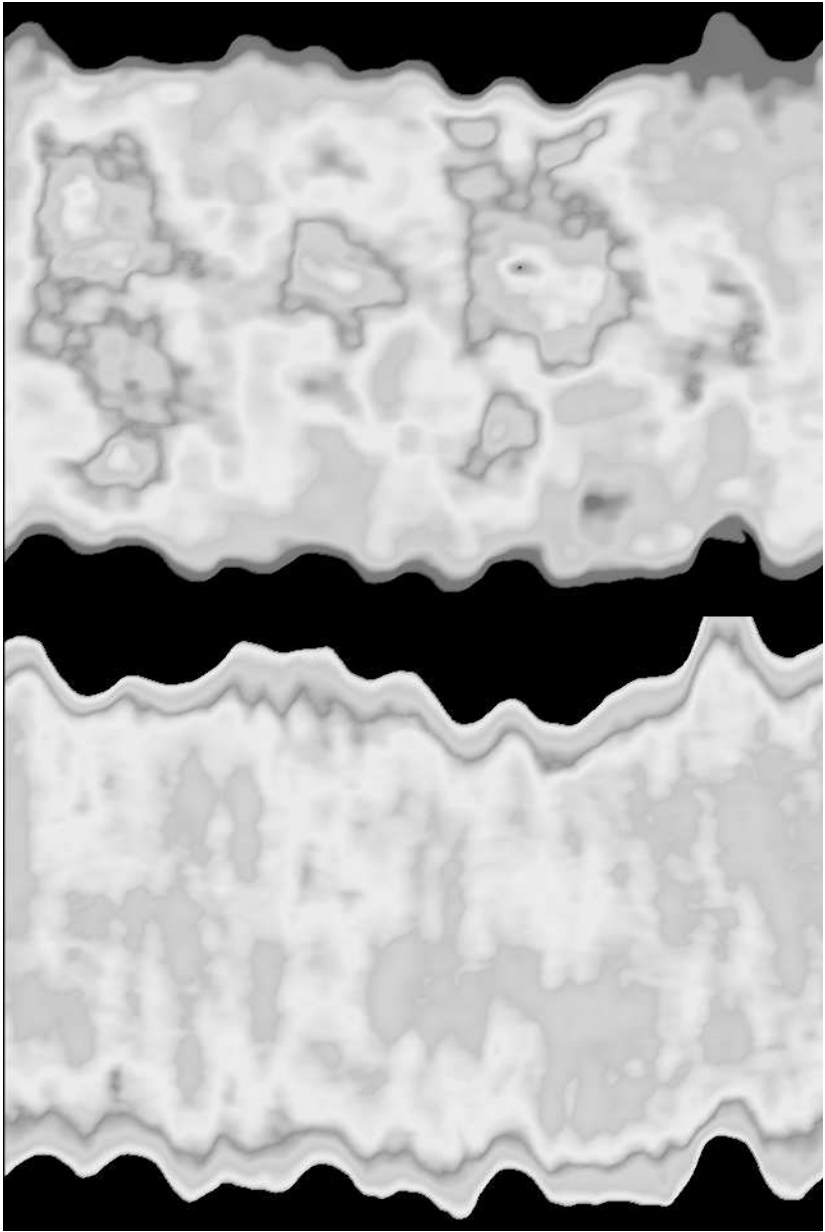}
\caption{(top) The density structure in an $800\times640$ simulation,
showing the inner $480\times640$ rectangle. The initial magnetic
field is vertical and the excitation zones are off the picture, to the
top and bottom of the grid. The density is color coded to emphasize the
hierarchical structure, with two cycles of color going from blue
to red as the density increases. The peak density is 1.826, 
the black level at the edge is 1.23, and the
minimum density at this time step, which occurs outside the rectangle
shown, is 0.19. (bottom) The corresponding temperature structure, with
red hotter than blue, in two cycles.
The clumps are warmer than the
interclump medium, indicating that the clumps are not confined by
thermal pressure from the interclump medium, but by the kinematic
pressure from their motion through this interclump medium. NOTE: gray
only for astro-ph. }
\label{fig:hier} \end{figure}

\newpage
\begin{figure} 
\vspace{.5in}
\caption{MPEG movie of the $800\times640$ simulation shown in 
figure 5, with density color coded in three cycles from blue to 
red, with blue at densities of 0.3, 1.7, and 2.0.}
\label{fig:mpeg}
\end{figure}

\newpage
\begin{figure}
\vspace{6.in}
\includegraphics{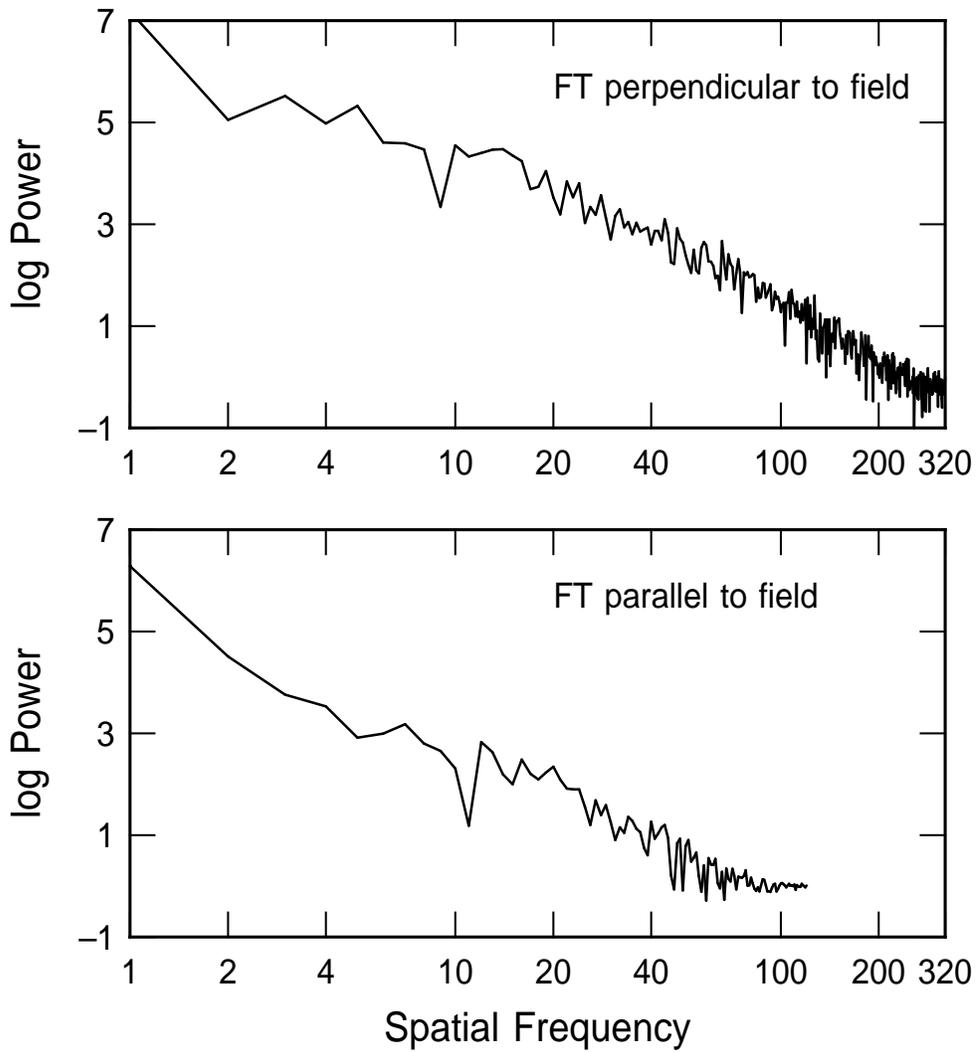}
\caption{Power spectra of the density structure for the model shown in
figure 5. The average power spectrum for the direction parallel to the
field is shown at the bottom, and for the direction perpendicular to the
field at the top. } \label{fig:fft} \end{figure}

\newpage
\begin{figure}
\vspace{6.in}
\includegraphics{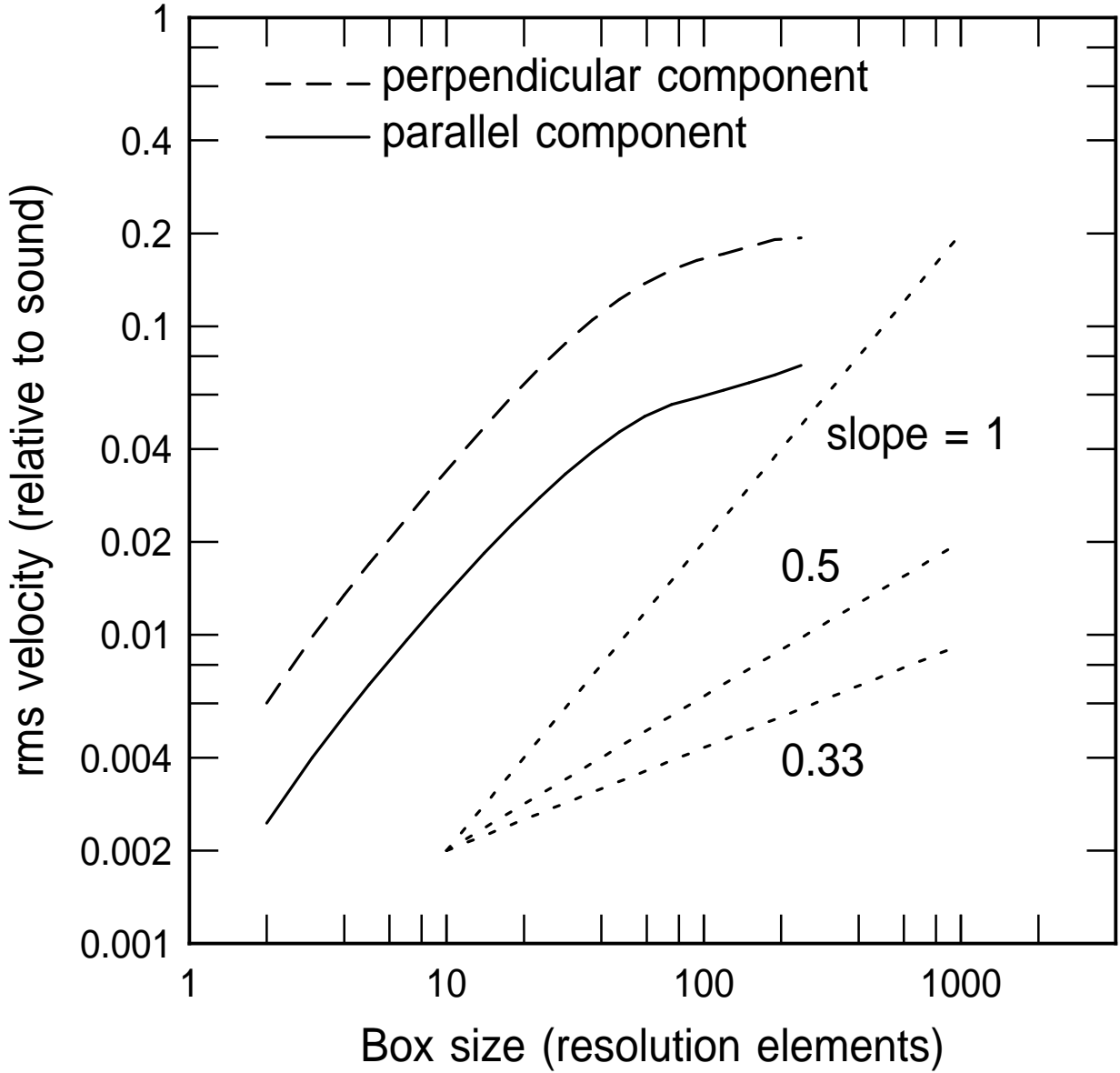}
\caption{The rms velocity in boxes of various sizes is shown versus the
box size for the simulation in figure 5. This diagram indicates
that the motions in the simulation are correlated with a velocity-size
relation given by a power law with power $\sim1/3$ for scales larger
than $\sim30$ cells, and with steeper power, $\sim1$ on smaller scales.
The grid boundaries are on much larger scales, 640 and 800, so edge
effects are not important here. } \label{fig:vel} \end{figure}

\newpage
\begin{figure}
\vspace{6.in}
\includegraphics{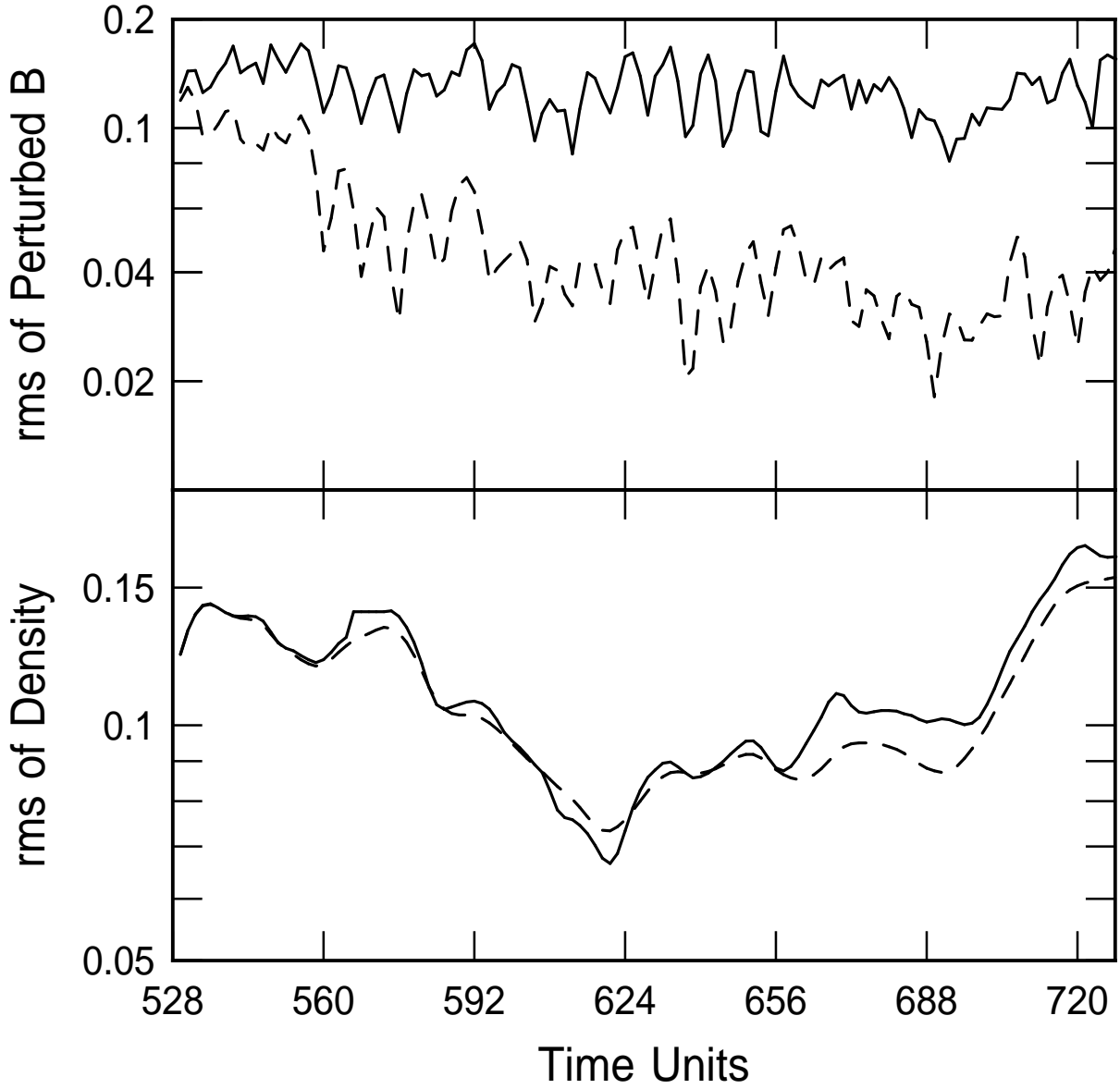}
\caption{The rms deviations around the mean, measured perpendicular to
the initial magnetic field, for density at the bottom of the figure and
perpendicular component of the field strength at the top, for two runs
with different magnetic diffusion rates. The dashed line has more
magnetic diffusion by a factor of 10 than the solid line. Enhanced
magnetic diffusion decreases the strength of the magnetic waves inside
the cloud without affecting the density structure. } \label{fig:rmsbrho}
\end{figure}

\newpage
\begin{figure}
\vspace{6.in}
\includegraphics{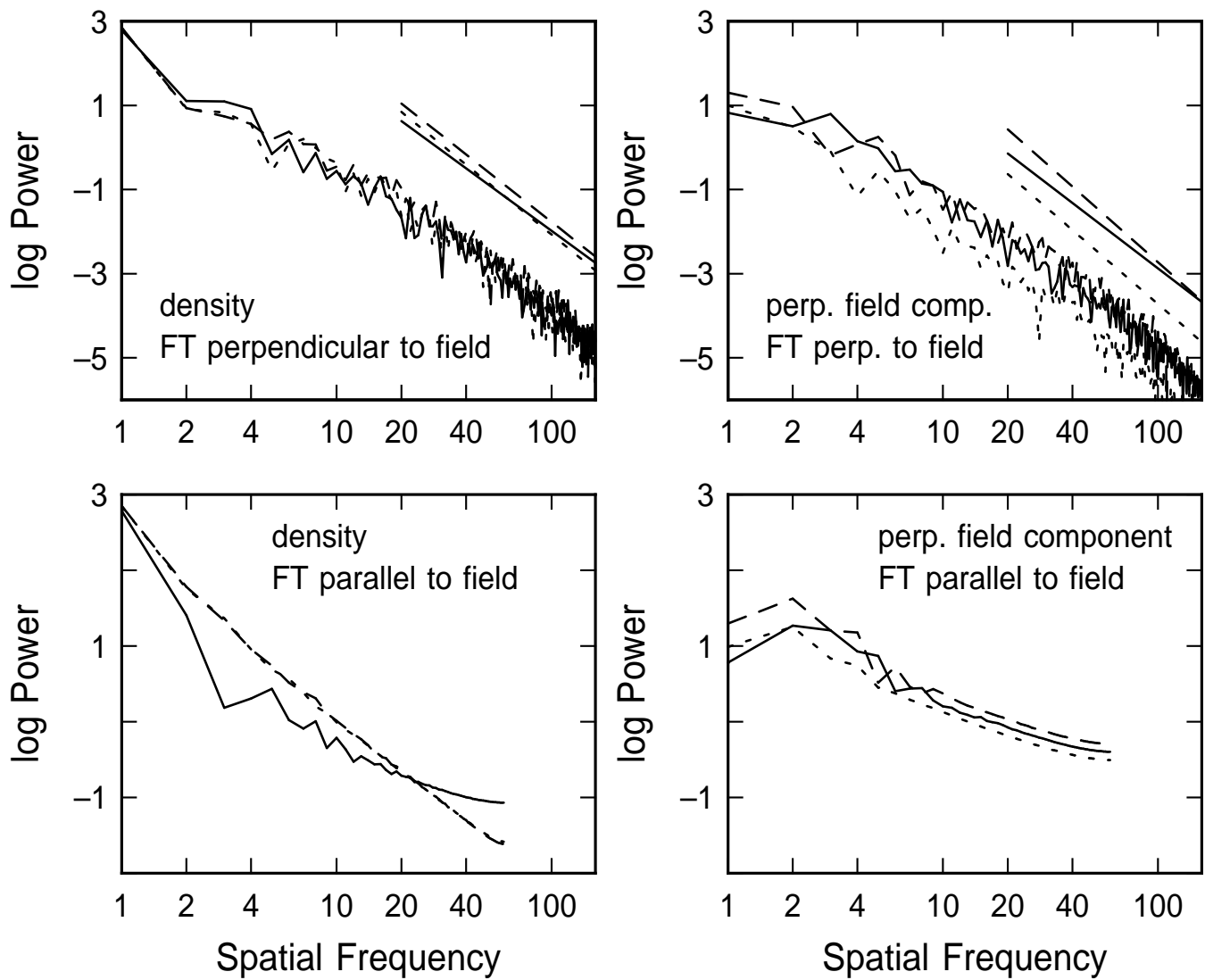}

\caption{The power spectra of density on the left and perpendicular
component of the magnetic field strength on the right, measured in
directions parallel and perpendicular to the initial field in the bottom
and top. The three line types correspond to the beginnings and ends of
two simulations: The solid curves correspond to times when the two
simulations have just completed identical initial cloud formation
phases. The dashed curve is at the end of one of the simulations,
following a continuation of the same slow magnetic diffusion. The dotted
curve is at the end of the other simulation, which had 10 times faster
magnetic diffusion during the last part. The straight lines with the
same line types show least squares solutions over the indicated ranges
of spatial frequency, shifted upwards by a factor of 100 for clarity.
Only the wave amplitude measured by the top right diagram shows any
significant change with the enhanced diffusion. The density structure is
relatively unchanged when the diffusion rate increases.}
\label{fig:fftbrho} \end{figure}

\newpage
\begin{figure}
\vspace{6.in}
\includegraphics{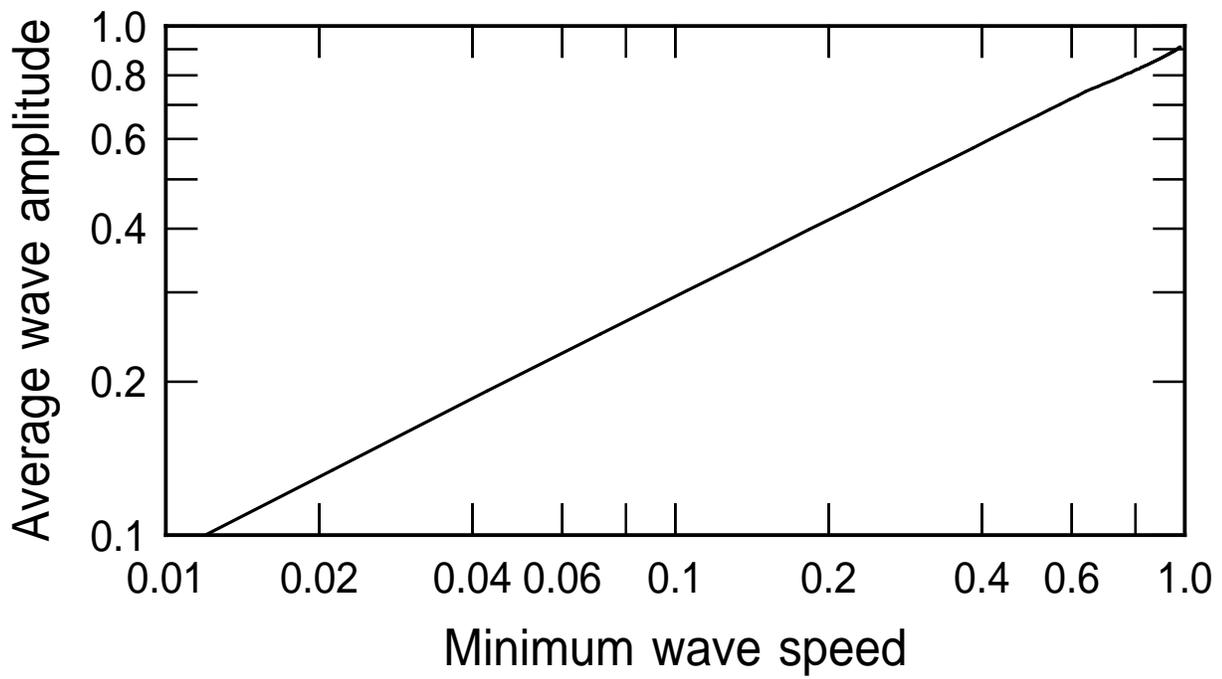}

\caption{The average wave amplitude inside one scale height is shown
versus the minimum wave speed at the center of the coordinate system for
a WKB solution of a wave propagating into a region with a central
depression in the wave speed.} \label{fig:wkb} \end{figure}

\newpage
\begin{figure}
\vspace{6.in}
\includegraphics{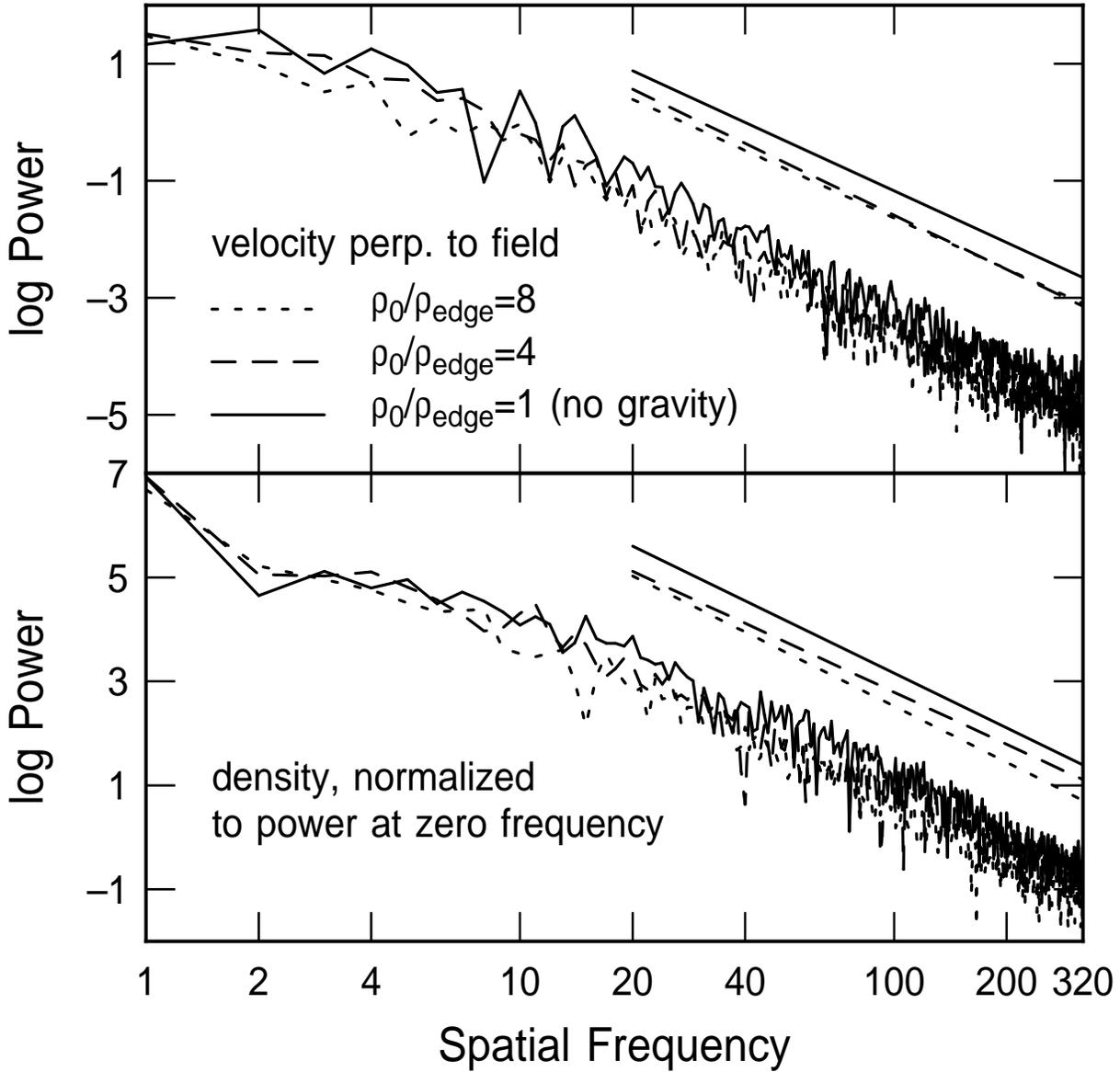}
\caption{Power spectra of density, in the bottom diagram, and
perpendicular velocity in the top, measured perpendicular to the field,
for three simulations with varying amounts of central density
concentration produced by fixed, plane-parallel, gravitational fields.
The density power spectrum is normalized to the power at 0 spatial
frequency. The solid curve is the same simulation shown in figure 5,
with no gravity and an initially uniform density. The dashed curve has
an initial equilibrium density enhancement in the center of the grid
that is a factor of 4 over the density at the edge, where the waves are
generated. The dotted curve has a factor of 8 equilibrium density
enhancement. The straight lines are least-squares fits to the power
spectra, shifted upwards by factors of 100 for clarity. The simulations
with enhanced central densities have weaker waves and smoother density
structures than the initially uniform solution, as shown by the smaller
power in the dashed and dotted lines compared to the solid lines. }
\label{fig:grav} \end{figure}

\end{document}